\documentstyle[12pt]{article}

\newcommand{\sect}[1]{\setcounter{equation}{0}\section{#1}}

\newtheorem{theorem}{Theorem}[section]
\newtheorem{proposition}[theorem]{Proposition}

\newtheorem{corollary}[theorem]{Corollary}
\newtheorem{lemma}[theorem]{Lemma}

\def\fraz#1#2{{\strut\displaystyle #1\over\displaystyle #2}}

\def\dim {{\sl Proof.}\phantom{X}}
\def\fidi{\hskip5pt \vrule height4pt width4pt depth0pt}
\def\hk{{\cal H}_N(\kappa)}

\def\fin{\fidi}

\def\N{{\bf N}}
\def\Z{{\bf Z}}
\def\R{{\bf R}}
\def\C{{\bf C}}

\def\T{{\bf T}}

\def\a{{\bf a}}

\def\tor{{{\T}^2}}
\def\eff{{\cal F}}

\def\hn{{\cal H}_{N}(\kappa)}
\def\hh{{\cal H}_{\hbar}(\kappa)}

\def\e#1{e^{\displaystyle #1}}
\def\opw#1{{Op^W(#1)}}
\def\opwo#1{{Op^W_0(#1)}}
\def\opaw#1{{Op^{AW}_{z}(#1)}}
\def\opwk#1{{Op^W_\kappa(#1)}}
\def\opawk#1{{Op^{AW}_{\kappa, z}(#1)}}
\def\slr{{{\rm SL}_2(\R)}}
\def\slz{{{\rm SL}_2(\Z)}}
\def\l#1{{\check{\ell}_{1,#1}}}

\def\ma#1{{{\rm M(A)}^{#1}}}
\def\mak#1{{{\rm M_\kappa(A)}^{#1}}}
\def\U#1#2{{{\rm U}({#1},{#2})}}
\def\norm#1{{\left\|{#1}\right\|}}

\begin{document}

\title{Exponential mixing and $|\ln\hbar|$ time scales \\
in \\
quantized hyperbolic maps on the torus}
\author{Francesco Bonechi\\
INFN, Sezione di Firenze \\Dipartimento di Fisica, Universit\`a di
Firenze \\ Largo E.Fermi 2, 50125 Firenze, Italy\\ e-mail:
bonechi@fi.infn.it\\ and\\ Stephan De Bi\`evre\\ UFR de
Math\'ematiques et UMR AGAT\\Universit\'e des Sciences et
Technologies de Lille\\ 59655 Villeneuve d'Ascq Cedex
France\\e-mail: debievre@gat.univ-lille1.fr }
\date{\today}

\maketitle
\begin{abstract}
{We study the behaviour, in the simultaneous limits $\hbar\to 0, t\to
\infty$,
of the Husimi and Wigner distributions of initial coherent states and
position eigenstates,  evolved under the
quantized hyperbolic toral automorphisms
and the quantized baker map. We show how the exponential mixing of the
underlying
dynamics manifests itself in those quantities on time scales logarithmic in
$\hbar$. The phase space distributions of the coherent states, evolved under
either of those dynamics, are shown to equidistribute on the torus in the
limit $\hbar \to 0$, for times $t$ between $\frac{1}{2}\frac{|\ln
\hbar|}{\gamma}$ and
$\frac{|\ln \hbar|}{\gamma}$, where $\gamma$ is the Lyapounov exponent of
the classical system.   For times shorter than $\frac{1}{2}\frac{|\ln
\hbar|}{\gamma}$, they
remain concentrated on the  classical trajectory of the center of the coherent
state.
The behaviour of the phase space distributions of evolved position
eigenstates, on the other hand, is not the same for the quantized
automorphisms as for the baker map.
In the first case, they equidistribute provided $t\to\infty$ as $\hbar \to 0$,
and as long as $t$ is shorter than $\frac{|\ln \hbar|}{\gamma}$. In the second
case, they remain localized on the evolved initial position at all such
times.}
\end{abstract}
\thispagestyle{empty}

\sect{Introduction}
It has been known for a long time  and proven in a large number of situations
that the eigenfunctions of a quantum system that has an ergodic classical
limit equidistribute on the relevant energy surface as $\hbar \to 0$
\cite{bo, bodb1, bodb2,
 cdv, dbde, gl, hmr, sc, z1}.
Similarly, if the underlying dynamics is mixing, this
 has an effect on the off-diagonal
matrix elements of observables between eigenstates \cite{bo, cr1, z2, z3}.
In other words, signatures
of ergodicity or mixing on {\em spectral} properties of quantum systems have
been
extensively studied. In this paper we exhibit a phenomenon in the {\em time}
domain
that is a signature of the {\em exponential} mixing of the underlying
dynamics: the
equidistribution on  the relevant energy surface of the Husimi and Wigner
distributions of an evolved coherent state in the limit where simultaneously
$\hbar\to0$  and $\frac{1}{2}\frac{|\ln\hbar|}{\gamma}<<t<<
\frac{|\ln\hbar|}{\gamma}$. We show this phenomenon
occurs in the quantized cat and Baker maps. We expect it to hold also in
other
exponentially mixing systems, such as perturbed cat maps, chaotic billards,
or manifolds of constant negative curvature. Rigorous proofs of these latter
statements are still lacking for reasons that will be explained below.

More precisely,  we give a rigorous proof of  the occurrence  of two distinct
time  scales, both logarithmic in $\hbar$, in the behaviour of the Wigner and
Husimi phase  space distributions of initial coherent states evolved under
quantized hyperbolic  toral automorphisms or the quantized baker map.
(Theorems \ref{cat} and \ref{baker}).  Denoting by $\gamma$ the Lyapounov
exponent of the classical map, we show in particular a qualitative difference
between the behaviour of the above quantities for times shorter than
$\frac{|\ln\hbar|}{2\gamma}$ and for times between
$\frac{|\ln\hbar|}{2\gamma}$ and $\frac{|\ln\hbar|}{\gamma}$. When
$t<<\frac{|\ln\hbar|}{2\gamma}$ the system is in what could be called,
following \cite{cach}, the ``orbital instability regime": the phase space
distribution functions behave as a delta function situated on
the classically evolved center of the initial coherent state. For times
$\frac{|\ln\hbar|}{2\gamma}<<t<<\frac{|\ln\hbar|}{\gamma}$ the system is in
the
``mixing" or ``statistical relaxation" regime: the phase space density
spreads
throughout the phase space and approaches the steady state given here by the
uniform distribution. Some numerical evidence for the existence of such a
transition at times of order $|\ln\hbar|$ was given in \cite{toik} for
the cat map and in \cite{bavo} for the Baker transformation without however
pinning down its precise location.

The occurence of the transition at $\frac{|\ln\hbar|}{2\gamma}$ is linked to
the support properties of the phase space distributions of coherent states
and
the situation is different for position eigenstates. When evolved under
a quantized toral automorphism, the latter reach the mixing regime
``immediately";  by this we mean in any limit, however slow, where $t\to
\infty$ as $\hbar\to0$, and for all times
up to $\frac{|\ln\hbar|}{\gamma}$. When evolved under the quantized baker
map, on the contrary, they remain in the ``orbital instability regime" for
all such times. For precise statements, see (\ref{posmix}) and
Proposition \ref{q_baker_position} (i).

We will argue below that the above phenomena can be understood in terms of
the interaction between the exponential mixing properties of the classical
dynamics and the support properties of the initial phase space distributions.
This suggests that the results should remain valid -- mutatis mutandis -- for
other exponentially unstable chaotic systems. To prove the existence of a
transition in the behaviour of the coherent state matrix elements, however,
one would need good control on the semi-classical approximation of the full
quantum evolution of such
systems for times longer than $|\ln \hbar|/2\gamma$, where $\gamma$ is the
classical mixing rate of the system. This is known to be a delicate matter:
see \cite{br, bgp, cr2, hj} for results in this direction.

The study of time-dependent quantities has attracted  less attention in
quantum chaos than that of spectral properties. Nevertheless, its importance
has
been stressed by several authors very early on \cite{b} \cite{beba}
\cite{bebatavo}. There
are two reasons for this. The first is that resolving fine spectral details
is
equivalent, via the Fourier transform, to understanding the (very) long time
behaviour of the system. The other one, perhaps a bit more philosophical, is
that
chaotic properties of classical dynamical systems (ergodicity, mixing and
mixing
rates, Lyapounov exponents,
\dots) manifest themselves asymptotically in time ($t\to \infty$). It is well
kown that bound quantum
mechanical systems, having discrete spectrum, can not be
chaotic in the classical sense. This poses the question of how classical
chaos
``emerges" from quantum mechanics as $\hbar\to 0$ and hence of the
non-commutativity of the $\hbar\to 0$ and $t\to \infty$ limits. The following
view on this problem has been expressed more or less explicitly by several
authors \cite{bavo}\cite{octo}\cite{cach}. We know that for times that are
not too
long, but that could still go to infinity as  $\hbar$ goes to $0$, classical
and
quantum mechanics are close. It is reasonable to expect that if, within this
time
scale, the classical system exhibits certain chaotic properties, then those
will
also
manifest themselves in the quantum system. Our results give a precise
quantitative content to this idea by showing that, for the quantized
cat and Baker maps, the classical exponential mixing manifests itself already
in the quantum system for times  shorter than $\frac{|\ln\hbar|}{\gamma}$.

Let us now give a precise statement of our results. For more details on the
objects introduced, we refer to section \ref{rev_mod} for the cat maps and to
section \ref{qbaker1} for the baker map. Let
$\hk (\kappa
\in [0,2\pi[\times [0,2\pi[)$ be the quantum Hilbert space of states
associated
to the torus $\tor$, where $N=1/(2\pi\hbar)\in\N$,  and where $\opwk{f}$ is
the
Weyl
quantization of the classical observable $f\in C^\infty(\tor)$. We also
introduce, for each
$x=(q, p)\in\tor$, the coherent state $|x,z,\kappa\rangle\in\hk$ which is
obtained by periodizing the Gaussian centered at $x$, with shape determined
by
$z\in\C$ ($Im\,z>0$). To each hyperbolic toral automorphism $A\in\slz \,
(|\hbox{Tr}A|>2)$ the metaplectic representation associates a unitary map
$M_\kappa(A)$ on $\hk$, the so called quantum map. To each (bounded) function
$f$ on the torus we associate the following ($t,N$) dependent function
$(x_0\in\tor)$:
\begin{equation}
\label{observable_t}
O_f^{AW,\kappa}[t,N](x_0) =
\langle x_0,z,\kappa | \mak{-t}\opawk{f}\mak{t} |x_0,z, \kappa \rangle\;.
\end{equation}
Here $\opawk{f}$ stands for the anti-Wick quantization of $f$,
so that
\begin{equation}
\label{aw}
O_f^{AW,\kappa}[t,N](x_0) =
\int_\tor f(x) |\langle x,z,\kappa|\mak{t} |x_0,z, \kappa \rangle|^2 \
\frac{dx}{2\pi\hbar}\;.
\end{equation}
Our principal result for the quantized hyperbolic toral automorphisms is
then:
\begin{theorem} \label{cat}For all $f\in C^\infty(\tor)$, for all $\epsilon
>0$
\begin{equation}
\label{classical_limit_intro}
\lim_{\stackrel{N\rightarrow\infty}{t<(1-\epsilon)\frac{\ln N}{2\gamma} }}
\left(O_f^{AW, \kappa}[t,N](x_0)-f(A^tx_0)\right) = 0,
\end{equation}
 and
\begin{equation}
 \label{mixing_limit_intro}
\lim_{\stackrel{N\rightarrow\infty}{(1+\epsilon)\frac{\ln N}{2\gamma}< t
< (1-\epsilon)\frac{\ln N}{\gamma}}}
O_{f}^{AW,\kappa}[t,N](x_0) = \int_\tor f(x) \,dx  \;,
\end{equation}
uniformly in $x_0\in\tor$ and in $t$ in the indicated region.
 Here $\gamma$ denotes the Lyapounov exponent of the classical map.
\end{theorem}
The first part of Theorem \ref{cat} asserts that if $N$ goes to
infinity with $t$ ``much" smaller than $\frac{\ln N}{2 \gamma}$, then
$O_f^{AW, \kappa}[t,N](x_0)$ tends to
$f(A^tx_0)$. In other words, on this time scale, the behaviour  of $O_f^{AW,
\kappa}$ is determined by the motion of the center of the coherent state,
which
 is erratic since  the classical map is chaotic  and which therefore depends
strongly on the initial $x_0\in\tor$ and on $A$:
the system is in the orbital instability regime. Note that,
in view of (\ref{aw}), equation
 (\ref{classical_limit_intro}) means that the Husimi distribution of the
evolved coherent states converges (weakly) to the delta function at $A^tx_0$
as
$\hbar$ tends to $0$.  Changing from  anti-Wick quantization to Weyl
quantization, it is easy to see that the same statement holds for the Wigner
distribution of the evolved coherent state as well.

Equation (\ref{mixing_limit_intro}) shows that, at times beyond $\ln
N/2\gamma$,
the situation changes.
In the asymptotic region where $\ln N/2\gamma<< t<< \ln N/\gamma$,
$O_f^{AW, \kappa}[t,N](x_0)$ becomes a constant,
independent of the coherent state center $x_0$ and of the dynamics $A$.
This result is a consequence of
the exponential mixing property of the classical map as we  explain below.

Equation (\ref{classical_limit_intro}) is of course easily understood
intuitively
using the
following simple argument about the evolution of coherent states. Recall
first
that (see also section 2) $M_\kappa(A)|x_0,z, \kappa\rangle
=(\hbox{phase})|Ax_0,A\cdot z, \kappa\rangle$, where the change from $z$ to
$A\cdot
z$ represents the stretching and squeezing of the initial coherent state
under
the classical evolution by a factor $\exp\gamma$.   Hence, if
$t\ll\frac{\ln N}{2\gamma}$ (i.e. $t\leq (\frac{1}{2}-\epsilon)\frac{\ln
N}{\gamma}$), the region in which the coherent state lives is of maximal
linear
size
$e^{\gamma t}/\sqrt{N}\approx N^{-\epsilon}$, which shrinks with $N$, even
though $t$ may go to infinity. As a result, the only contribution to the
integral in calculating  (\ref{aw})
comes from the point $A^tx_0$.

To understand (\ref{mixing_limit_intro}), let
us first recall that for the classical system one has (see \cite{db} for a
simple proof of this well known fact):
\begin{equation}
\label{expmix}
|\int_{\tor} f(x) \rho(A^{-t}x, t) \ dx - \int_{\tor} f(x)\ dx|\leq
C_f\left(\int_{\tor} |\nabla\rho|(x, t) \ dx \right)\exp-\gamma t,
\end{equation}
for any phase space distribution $\rho(x, t) \geq0, \ \int_\tor \rho(x, t) \
dx =1$, which may depend explicitly on $t$, a remark we will use below. This
means $\rho(x, t)$ converges (weakly) to $1$ and does so exponentially fast
provided $\sup_t\int_{\tor} |\nabla\rho|(x, t) \ dx<\infty$: this is the
property called exponential mixing, a consequence of the exponential
instability of the dynamics. Note however that the convergence depends also
on
$\int_{\tor} |\nabla\rho|(x, t) \ dx$, a quantity that should be thought of
as
measuring the scale on which $\rho(x, t)$ fluctuates: the faster the
fluctuations
in $\rho$, the bigger this quantity is and the slower  the convergence in
(\ref{expmix}). Consider now
$$
O_f^{AW, \kappa}[t,N](x_0)= \int_\tor f(x) \frac{|\langle x_0, z,
\kappa|A^{-t}x,
A^{-t}\cdot z, \kappa\rangle|^2}{2\pi\hbar}\ dx.
$$
This is precisely of the form $\int_{\tor} f(x) \rho(A^{-t}x, t) \ dx$ with
$$
\rho(x, t) = \frac{|\langle x_0, z, \kappa|x,
A^{-t}\cdot z, \kappa\rangle|^2}{2\pi\hbar}.
$$
Hence $\rho(x, t)$ is the Husimi distribution of the initial coherent state
$|x_0,z,\kappa\rangle$ with respect to the ``squeezed" coherent states $|x,
z'=A^{-t}\cdot z,\kappa\rangle$: as a result, it is not unreasonable to
expect
 that $\int_{\tor}|\nabla\rho|(x, t) \ dx$ is of order $1/\sqrt{\hbar}$.
If this were true, the exponential mixing property of the classical dynamics
would immediately imply that
$$
|O_{ f}^{AW,\kappa}[t,N](x_0) - \int_\tor f(x) \,dx|\leq
C_f\frac{\exp-\gamma t}{\sqrt\hbar},
$$
yielding (\ref{mixing_limit_intro}).  We will give two proofs of
(\ref{mixing_limit_intro}). The proof of  Proposition \ref{direct_proof} (ii)
will
follow essentially the above strategy which clearly brings out the role of
the
exponential mixing in the result, and its interaction with the natural length
scale $\sqrt\hbar$ associated to the Husimi distribution of
$|x_0,z,\kappa\rangle$. The proof of Proposition
\ref{plane_limit}-\ref{torus_limit}
is simpler, but perhaps less telling. Let us also point out that
(\ref{mixing_limit_intro}) remains true, even if $x_0$ is a periodic point of
the
underlying dynamics.

The toral automorphisms are obtained by folding a linear dynamics back
over the torus and
as a result they enjoy, both classically and quantum-mechanically, very
special
arithmetic properties \cite{degris, karu, ke}. This makes them
rather particular and it is of interest to see
if a result similar to Theorem \ref{cat} holds for more
 ``generic" chaotic systems, such as the perturbed cat maps or the baker map.
This would ensure that the phenomena observed are not due to the special
nature
of the toral automorphisms.  A partial answer to this question is given
for the baker map in Theorem \ref{baker}
below. Our principal tool in the proof of this result is an Egorov theorem
(Proposition \ref{egorov_baker}) proven for
this system in \cite{dbde}, which allows us to control the dynamics for times
up to
$\log_2N$. Recall that the Lyapounov exponent of the Baker map is $\ln2$. As
mentioned previously, such control is not available for
perturbed
toral automorphisms, nor for any other hyperbolic Hamiltonian system. Writing
$B$
for the baker map and $V_B$ for its quantization on
${\cal H}_N(0)$, we have (for unexplained notations, see section
\ref{qbaker}):
\begin{theorem}\label{baker}
$\forall x_0\in\tor, \forall 0<\epsilon<1$, there exists a sequence
$x_N=(q_N,p_N)\in \tor, ||x_N-x_0||\leq N^{-\epsilon/5}$, so that, for all
$f\in
C^\infty(\T)$,
$$
\lim_{\stackrel{N\rightarrow\infty}{0\leq t <
(1-\epsilon)\frac{1}{2}\log_2 N }} \langle x_N, z, 0| V_B^{-t}
Op_{0,z}^{AW}(f) V_B^{t}| x_N, z, 0\rangle
 - f(2^t q_N)
 =0
$$
and
$$
\lim_{\stackrel{N\rightarrow\infty}{(1+\epsilon)\frac{1}{2}\log_2
N < t < (1-\epsilon)\log_2 N }} \langle x_N, z, 0| V_B^{-t}
Op_{0,z}^{AW}(f) V_B^{t}| x_N, z, 0\rangle
 = \int_0^1 f(q) \,dq  \;,
$$
The limits are uniform for $t$ in the indicated region.
\end{theorem}

\noindent Comparing to Theorem \ref{cat}, this result has two
obvious shortcomings:
the restriction to functions $f$ of the $q$ variable alone and the use of
the sequences $x_N$.  Their origin will be explained in section
\ref{qbaker2}.

The occurence of a transition at $\frac{|\ln\hbar|}{2\gamma}$
in the qualitative behaviour of the matrix elements
above is directly related to the choice of a coherent state as an initial
state.
To illustrate this, we also study  the behaviour of
$\langle e_j^\kappa, M_\kappa(A)^{-t}\opawk f M_\kappa(A)^t
e_j^\kappa\rangle$
(section \ref{sec_catpos}),
and  of
$\langle e_j^\kappa, V_B^{-t}\opawk f V_B^t e_j^\kappa\rangle$
(section \ref{qbaker}),
where the $e_j^\kappa$ are the ``position eigenstates" in $\hh$
(see section \ref{rev_mod}).
In the case of the cat maps, if one applied the same heuristic reasoning as
above (based on (\ref{expmix})), one would conclude that
the mixing regime should set in {\em no later than} for times of the order
$\frac{\ln N}{2\gamma}$, since the Husimi distribution of the position
eigenstates
still has a spread of the order of $\sqrt \hbar$ in the $q$-direction.
In fact, we will prove in section \ref{sec_catpos} that for all sequences
$j_N (1\leq j_N \leq N)$
\begin{equation}\label{posmix}
\lim_{\stackrel{N\rightarrow\infty, t\to \infty}{ t
< (1-\epsilon)\frac{\ln N}{\gamma}}}
\langle e_{j_N}^\kappa, M_\kappa(A)^{-t}\opawk f M_\kappa(A)^t
e_{j_N}^\kappa\rangle
=\int_\tor f(x) dx,
\end{equation}
the limit being again uniform with respect to $j_N$: in other words, no
orbital
instability regime is observed here, mixing sets in as soon as $t\to\infty$.
 This results, intuitively at least,
from a suitably adapted  version of (\ref{expmix}), proven in section
\ref{qbaker},
together with the observation that the phase space distributions of the
position
eigenstates are completely delocalized in the $p$-direction.

In the case of the baker map, the opposite phenomenon occurs: in this case,
no mixing regime is observed for $\langle e_j^\kappa, V_B^{-t}\opawk f V_B^t
e_j^\kappa\rangle$ at times shorter than $|\log_2\hbar|$. For a detailed
statement we refer to Proposition \ref{q_baker_position} (i). This result can
again be understood in terms of the interaction of the mixing properties of
the  classical map with the support properties of the initial position
state, the support of which is now exactly lined up with the stable manifold
of the classical baker map.

As mentioned above, manifestations of classical chaos in the corresponding
quantum
system have been searched mainly in the energy domain. Nevertheless, several
papers \cite{b, beba, bebatavo, bavo, cach, octo, octohe, tohe, toik} have
analysed
the behaviour of
the Wigner function of semi-classically evolved
initial states with initial support either in a point, such as coherent
states, or
on a Lagrangian submanifold, such as position eigenstates, and compared them
numerically
to the Wigner function of the quantum mechanically evolved states. The
general
picture that emerges from these studies is as follows: for times that are not
too
long
the semi-classical approximation should be valid and is itself determined
essentially by
the behaviour of the support of the Wigner function of the initial state
under the
classical flow.  The results of the present paper, as summarized above,
corroborate this picture by proving some rigorous statements in this direction
on a few simple systems, for times up to $|\ln\hbar|/\gamma$. As an example,
Theorem \ref{cat} shows that the Husimi distribution of the
{\em quantum mechanically evolved} coherent state ({\em i.e.}
$\frac{|\langle x_0, z, \kappa|A^{-t}x,
A^{-t}\cdot z, \kappa\rangle|^2}{2\pi\hbar}$) can not be distinguished, as
$\hbar$ goes to $0$, from the {\em classically evolved} Husimi distribution
of the initial coherent state ({\em i.e.} , $\frac{|\langle x_0, z,
\kappa|A^{-t}x, z, \kappa\rangle|^2}{2\pi\hbar}$) for times up to $\frac{\ln
N}{\gamma}$.  The change in behaviour of either of these quantities,
 at times of the orderof $\frac{\ln N}{2\gamma}$ is entirely due to the
interaction of the exponential mixing of the classical dynamical system with
the uncertainty principle which forces the initial  Husimi distribution to
have a spread of size $\sqrt{\hbar}$.

In addition, we exhibit one extra phenomenon: a slightly longer time scale
on which the above picture breaks down, in the sense that the semi-classical
or quantum evolution does no longer stay close to the underlying purely
classical evolution. Indeed, we show in section \ref{sec_period} that,  in
quantized cat maps,
the classical-quantum (and hence the classical-semi-classical)
agreement  breaks down ``a little later than at $|\ln\hbar|/\gamma$", that is
to say at times of order  $\frac{3}{2}\frac{\ln N}{\gamma}$. More details, as
well as an intuitive explanation of the phenomenon can be found in section
\ref{sec_period}.

It is important to add a few words of caution: we want to stress
that our results have nothing to say about the problem of the
existence or not of a so-called $|\ln\hbar|$-barrier, {\em i.e.} a
time scale of the form $\alpha\frac{|\ln \hbar|}{\gamma}\
(\alpha>0)$ beyond which semi-classical approximations to the
quantum evolution may break down \cite{b, beba, bebatavo, bavo,
octo, octohe, tohe}. What we have shown here is that the
exponential mixing of the classical dynamics manifests itself in
the quantum system already on time scales short enough to be
accessible to semi-classical approximations. In other words, the
phenomena we exhibit are taking place before such a breakdown --
if any -- is expected to occur. At any rate, the cat maps being
linear, for them semi-classical approximations are exact at all
times, and their study  can not shed light on the above problem.
For the baker map, we show that the semi-classical evolution of
the Husimi or Wigner distributions of a position eigenstate
(Proposition \ref{q_baker_position} (ii)) and of a coherent state
(Proposition \ref{q_baker_cs} (ii)) agrees indeed with the quantum
mechanical one for times up to $\log_2 \hbar$, as expected.

At the risk of belabouring the point, we repeat one more time
that we have not been able to exhibit a time scale on which the agreement
between the semi-classical and the full quantum evolution breaks down.

\sect{Kinematical estimates on the torus} \label{rev_mod}
The
kinematic framework of quantum mechanics on the two-torus $\tor$
(with coordinates $(q,p)\in[0,1[\times[0,1[$) as well the
quantization of the toral automorphisms is well known. We briefly
recall the essential ingredients, following \cite{bo, bodb2}, where proofs
omitted here, as well as references to the original literature can be found.

First, recall that the quantum Hilbert space of a particle on the line is
$L^2(\R)$.
Moreover, to each classical observable $f(q,p)\, (f\in C^2(\R^2))$,
Weyl quantization associates a quantum observable $\opw f$ defined by
\begin{equation}
\opw f= \int\int \tilde f(\a)
U(\a)\frac{d\a}{2\pi\hbar},
\hbox{where}
f(q,p) = \int \tilde f(\a)
\e{-\frac{i}{\hbar}(a_1p-a_2q)}\ \frac{d\a}{2\pi\hbar},
\end{equation}
and where
$
U(a_1,a_2) = \e{-\frac{i}{\hbar}(a_1P-a_2Q)}
$
are the phase space translation operators in the usual notations
(see \cite{bodb2} for
further technical details). In particular, if $f$ is periodic of period $1$
in both
$q$ and $p$, one readily concludes that
\begin{equation}
\opw f = \sum_{n\in\Z^2} f_{n} U(2\pi\hbar n_1, 2\pi\hbar n_2),\,
\hbox{ where } f=\sum_{n} f_{n} \e{2\pi i(n_2q-n_1p)}.\label{wqper}
\end{equation}
We will always suppose $\sum_n|f_n|<\infty$ and shall write
\begin{equation}
f\in\l{k} \, {\rm iff} \, \sum_{n}|f_{n}|\norm{n}^k <\infty.
\end{equation}
For $\xi,\eta\in\R^2$, we introduce the notation
$\langle\xi,\eta\rangle=\xi_1\eta_2-\xi_2\eta_1$ and
$\chi_\xi(x)=\exp 2\pi i\langle x,\xi\rangle$.

One constructs the quantum Hilbert space of states by searching
for states $\psi$ having the symmetry of the torus, i.e. states
periodic of period $1$ both in the position and the momentum
variable:
\begin{equation}U(1,0)\psi = \e{-i\kappa_1} \psi    ~~~~~~~~~
U(0,1)\psi = \e{i\kappa_2} \psi,
~(\kappa_1,\kappa_2)\in[0,2\pi)\times[0,2\pi),
\label{periodic}
\end{equation}
where one allows for the usual phase change.  The space of solutions
$\hh$ to (\ref{periodic})  is a space of tempered distributions of which one
proves
readily
 that either
it is zero-dimensional, or there exists a positive integer $N$ so that
\begin{equation}
2\pi\hbar N =1,\label{bs}
\end{equation}
in which case it is $N$-dimensional. Equation (\ref{bs}) will always be
assumed to hold in the rest of this section.

Condition (\ref{bs}) is equivalent to $\bigl[U(0,1),U(1,0)\bigr]=0$,
so that (\ref{periodic}) is nothing but
the problem of simultaneously diagonalizing $U(0,1)$ and $U(1,0)$.  Since
their
spectra are
continuous, this leads to a direct integral decomposition of $L^2(\R)$:
\begin{equation}\label{dirint}
L^{2}(\R)\cong
\int_0^{2\pi}\int_0^{2\pi}\frac{d\kappa}{(2\pi)^{2}}\ \hh\ ,\ \
\psi \cong \int_0^{2\pi}\int_0^{2\pi}\frac{d\kappa}{(2\pi)^{2}}\ \psi(\kappa).
\end{equation}
This equips each ``term" $\hh$ with a natural inner product. For
each admissible $\hbar$ (i.e. $2\pi\hbar N=1$), the spaces $\hh$,
indexed by $\kappa$, are the quantum Hilbert spaces of states for
systems having the torus as phase space. An orthonormal basis for
$\hh$ is given by the position eigenstates
$\{e^\kappa_j\}_{j=0}^{N-1}$, where
\begin{equation}
\label{base_pos}
e^\kappa_j(y) = \sqrt{\frac{1}{N}} \sum_n \e{in\kappa_1}
\delta(y-\frac{\kappa_2}{2\pi N} -\frac{j}{N}-n)  ~~~~ y\in \R \;.
\end{equation}
The following unitary representation of the discrete phase space translations
\begin{equation}
\label{heisenberg}
U_\kappa\left(\frac{n_1}{N},\frac{n_2}{N}\right)\, e^\kappa_j =
e^{i\pi\frac{n_1n_2}{N}} e^{i(\kappa_2+2\pi j)\frac{n_2}{N}}
e^\kappa_{j+n_1} \;
\end{equation}
is obtained by restriction of $U(\frac{n_1}{N},\frac{n_2}{N})$ to
$\hh$. For each $\psi\in {\cal S}(\R)$, we have that
\begin{equation}
\label{develop_pos_eig} \langle e^\kappa_j,\psi(\kappa) \rangle =
\frac{1}{\sqrt{N}} \sum_{n\in\Z} e^{i\kappa_1n}
\psi(\frac{\kappa_2}{2\pi N}+\frac{j}{N}-n) \;.
\end{equation}

For later use, we introduce the vectors $\epsilon_j\in L^2(\R)$
that are  the direct sum (see (\ref{dirint})) of the position eigenstates
$e_j^\kappa\in\hh$.

\medskip
\begin{lemma}\label{lem_posvec}
Let $\epsilon_j\in L^2(\R)$, $j=0,\ldots,N-1$, be defined by
$\epsilon_j(x)=\sqrt{N}$ if $j/N\leq x < (j+1)/N$ and $0$ otherwise.
Then
$$
\epsilon_j = \int_\oplus \frac{d\kappa}{(2\pi)^2} \ e^\kappa_j \;.
$$
\end{lemma}

\noindent\dim A direct computation shows that for each $\psi=\int_\oplus
\frac{d\kappa}{(2\pi)^2} \
\psi(\kappa)$ one has
$
\langle \epsilon_j,\psi\rangle = \int d\kappa/(2\pi)^2 \ \langle
e^\kappa_j,\psi(\kappa)\rangle\;. \fidi
$

Turning now to the quantization of
observables, since for each $f$ on $\tor$, the commutator $\bigl[\opw f,U(k,
\ell)\bigr]=0$, it is
clear that, for each $\kappa$,
$\opw f\ \hh\subset\hh$ and hence
\begin{equation}
\opw f\cong\int_0^{2\pi}\int_0^{2\pi}\frac{d\kappa}{(2\pi)^{2}}\  \opwk f,
\label{dirintw}
\end{equation}
where $\opwk f$ denotes the restriction of $\opw f$ to $\hh$.
We conclude from (\ref{wqper}) and (\ref{bs}) that
$$
\opwk f = \sum_{n\in\Z^2} f_n\  U_\kappa \left(\frac{n_1}{N},
\frac{n_2}{N}\right).
$$

We will make extensive use of coherent states in our analysis.
Recall first the definition of the standard
Weyl-Heisenberg coherent states. For
$z\in\C$, and $Imz>0$, they are defined, for each  $x=(q,p)\in\R^2$, by
\begin{equation}
\eta_{0,z}(y)=\left(\frac{Imz}{\pi\hbar}\right)^\frac{1}{4}\e{\frac{i}{2\hbar
}zy^{2
}},\quad
\eta_{x,z}(y)=(U(q,p)\eta_{0,z})\, (y)\equiv \langle y\mid x,z\rangle
~y\in\R,
\label{cs}
\end{equation}
in the standard manner. We will systematically  use the bra-ket
notation of Dirac, i.e. $\eta_{x,z}(y)=\langle y\mid x,z\rangle$.
Furthermore we recall here the explicit formula for the scalar
product between coherent states ($x=(q,p)$, $x'=(q',p')$)
\begin{equation}
\label{scalar_cs} \langle x,z|x',z' \rangle = F(z,z') \
\e{\frac{i}{2\hbar} (qp'-pq')}
\e{-\frac{1}{2\hbar}(x-x',B(z,z')(x-x'))}\,,
\end{equation}
where, $F(z,z') = ({\rm Im}z\ {\rm
Im}z')^{1/4}/\sqrt{(z'-z^*)/(2i)}$, and
$$
B(z,z') = \frac{1}{z'-z^*}\left(
\begin{array}{cc}
iz^*z' & -\frac{i}{2}(z'+z^*) \cr -\frac{i}{2}(z'+z^*)
& i
\end{array} \right)
\;.
$$
If $z=z'$ then $B(z)=B(z,z)$ is a positive definite matrix; we
will denote by  $\beta_+(z)$ ($\beta_-(z)$) its greatest
(smallest) eigenvalue.

The following property that generalizes the resolution of the
identity for coherent states is valid for each
$\psi_1,\psi_2,\phi_1,\phi_2\in L^2(\R)$:
\begin{equation}
\label{gen_res_id} \int_{\R^2} \frac{dx}{2\pi\hbar}
\langle\psi_1|U(x)\phi_1\rangle \langle U(x)\phi_2|\psi_2\rangle =
\langle \psi_1|\psi_2\rangle \langle\phi_2|\phi_1\rangle \;.
\end{equation}

The coherent states $|x,z,\kappa\rangle $ are defined on the torus
implicitly by
\begin{equation}\label{dirintcs}
\mid x, z\rangle \cong
\int_0^{2\pi}\int_0^{2\pi}\frac{d\kappa}{(2\pi)^{2}} \mid x, z,
\kappa\rangle\;.
\end{equation}
They still satisfy an analogue of the generalized resolution of
the identity.
\begin{lemma}
\label{gen_res_id_tor} For $i=1,2$, let $\psi_{i,\kappa_0}\in{\cal
H }_N(\kappa_0)$, $\phi_i\in{\cal S}(\R)$. If we denote by\\ $\int
dk/(2\pi)^2 [U(x)\phi_i](\kappa)$ the integral decomposition of
$U(x)\phi_i$ for $x\in\tor$ , \\we have
$$\int_{\tor}
\frac{dx}{2\pi\hbar}
\langle\psi_{1,\kappa_0}|[U(x)\phi_1](\kappa_0)\rangle \langle
[U(x)\phi_2](\kappa_0)|\psi_{2,\kappa_0}\rangle =
\langle\psi_{1,\kappa_0}|\psi_{2,\kappa_0}\rangle \langle
\phi_2|\phi_1\rangle \;.$$
\end{lemma}
\dim Recall from \cite{bodb2} that the map
$$
S(\kappa)= (\sum_m \exp -i\kappa_2m \ U(0,m))(\sum_n\exp
i\kappa_1n\  U(n,0))
$$
defines a surjection of the space of  Schwartz functions ${\cal
S}(\R)$ onto $\hh\subset {\cal S}'(\R)$. Let $\phi, \psi\in {\cal
S}(\R)$ and write, as in (\ref{dirint}), $\phi \cong \int
\frac{d\kappa}{(2\pi)^2}\ \phi(\kappa), \psi \cong \int
\frac{d\kappa}{(2\pi)^2}\ \psi(\kappa)$.  Then, as a simple
calculation shows,
$$
\langle \phi(\kappa), \psi(\kappa)\rangle = \langle \phi,
S(\kappa)\psi\rangle = \frac{1}{N}\sum_{\ell \in \Z}
\bar\phi(x_\ell)\psi^{(\kappa_1)}(x_\ell),
$$
where $x_\ell = \frac{\kappa_2}{2\pi N} + \frac{\ell}{N}$ and
$$
\psi^{(\kappa_1)}(y) = \sum_{n\in \Z}\exp i\kappa_1 n \ \psi(y-n).
$$
As a simple inspection shows, the matrix element $\langle
\phi(\kappa), \psi(\kappa)\rangle$ is a smooth period function of
$\kappa$ and, as a result, is pointwise equal to the sum of its
Fourier series. These observations can be used to justify the
following formal computation. Let $\psi_i\in{\cal S}(\R)$ such
that $S(\kappa_0)\psi_i\cong\psi_{i,\kappa_0}$. Then we have
\begin{eqnarray}
\nonumber
& &\int_{\tor} \frac{dx}{2\pi\hbar}
\langle\psi_{1,\kappa_0}|[U(x)\phi_1](\kappa_0)\rangle \langle
[U(x)\phi_2](\kappa_0)|\psi_{2,\kappa_0}\rangle =\cr \nonumber
&~~& =\sum_{mn}e^{i\pi N(m_1m_2+n_1n_2)}
e^{i((m_1+n_1){\kappa_0}_1-(m_2+n_2){\kappa_0}_2)}
\phantom{\int_{\tor}} \cr
&~~& \quad\quad\quad\quad\quad\quad\quad\quad \int_{\tor}
\frac{dx}{2\pi\hbar}  \langle \psi_1|U(m)U(x)\phi_1\rangle \langle
U(x)\phi_2|U(n)\psi_2\rangle\cr \nonumber
&~~& =\sum_{ms}e^{i\pi
Ns_1s_2} e^{i (s_1{\kappa_0}_1-s_2{\kappa_0}_2)} \int_{\tor}
\frac{dx}{2\pi\hbar} \langle \psi_1|U(m+x)\phi_1\rangle \langle
U(x+m)\phi_2|U(s)\psi_2\rangle\cr \nonumber
&~~& =\sum_{s}e^{i\pi
Ns_1s_2} e^{i (s_1{\kappa_0}_1-s_2{\kappa_0}_2)} \int_{\R^2}
\frac{dx}{2\pi\hbar}  \langle \psi_1|U(x)\phi_1\rangle \langle
U(x)\phi_2|U(s)\psi_2\rangle\cr
&~~& =\sum_{s}e^{i\pi
Ns_1s_2} e^{i (s_1{\kappa_0}_1-s_2{\kappa_0}_2)} \langle
\psi_1|U(s)\psi_2\rangle \langle \phi_2|\phi_1\rangle =
\langle\psi_{1,\kappa_0}|\psi_{2,\kappa_0}\rangle \langle
\phi_2|\phi_1\rangle \;. \fidi
\end{eqnarray}

Given $f\in L^\infty(\tor)$, the anti-Wick quantization
$\opawk{f}$ of $f$ is the operator on $\hh$ defined by
\begin{equation}
Op^{AW}_{\kappa,z}(f) =\int_{\tor} f(x)\ |x,z,\kappa\rangle\langle
x, z, \kappa| \frac{dx}{2\pi\hbar}.
\end{equation}
One has \cite{bodb2},
\begin{equation}
 \hbox{Id}_{\hh}=\int_{\tor} \ |x,z,\kappa\rangle\langle x, z, \kappa|
\frac{dx}{2\pi\hbar};\, Op^{AW}_{z}(f) = \int^\oplus
\frac{d\kappa}{(2\pi)^2} \ Op^{AW}_{\kappa,z}{f}.
\end{equation}

In the following lemma we collect a useful relation between Weyl
and anti-Wick quantization.

\begin{lemma}
\label{weyl_antiW} Let $z\in\C$, with ${\rm Im}z>0$.

\noindent {\rm($i$) [Plane]} For each $\varphi_i\in L^2(\R), i=1,2$,
$\beta\in\R^2$ we have that
$$
\langle \varphi_1 | Op^{AW}_{z}(\chi_\beta)|\varphi_2\rangle =
\langle 2\pi\hbar\beta,z|0,z\rangle \langle
\varphi_1|\opw{\chi_\beta}|\varphi_2\rangle   \;.
$$
\noindent {\rm($ii$)[Torus]}
For each $\varphi_{i,\kappa}\in\hn$, $\beta\in\Z^2$ we have that
$$
\langle \varphi_{1,\kappa} |
Op^{AW}_{\kappa,z}(\chi_\beta)|\varphi_{2,\kappa}\rangle = \langle
\frac{\beta}{N},z|0,z\rangle \langle
\varphi_{1,\kappa}|\opwk{\chi_\beta}|\varphi_{2,\kappa}\rangle
\;.$$
\end{lemma}
\dim (i) The result comes from (\ref{gen_res_id}) with
$\phi_1=|0,z\rangle$, $\phi_2=U(2\pi\hbar\beta)|0,z\rangle$,
$\psi_1=\varphi_1$, $\psi_2=U(2\pi\hbar\beta)\varphi_2$. Point
(ii) is proven exactly as in (i) using now Lemma
(\ref{gen_res_id_tor}) with $\phi_1=|0,z\rangle$,
$\phi_2=U(\beta/N)|0,z\rangle$,
$\psi_{1,\kappa}=\varphi_{1,\kappa}$ and
$\psi_{2,\kappa}=U_\kappa(\beta/N)\varphi_{2,\kappa}$. \fidi

Let us introduce some notation. If $s\in\R$, then we denote by
$p(s)$ the nearest integer to $s$; if $s=m+1/2$ is a half-integer
($m\in\Z$) then $p(m+1/2)=m$. With some abuse of notation, we
write $p(\xi)=(p(\xi_1),p(\xi_2))\in\Z^2$ for $\xi\in\R^2$.

We end this section with two propositions collecting a few simple but crucial
semi-classical estimates on the Weyl quantized observables and the Husimi
distributions, respectively.
\begin{proposition}
\label{fourier_gen} Let $z\in\C$, with ${\rm Im}z>0$.
\noindent(i) For each function $f\in \l{0}$, for all $x=(q,p),
x'=(q',p')\in\tor$, we have that
\begin{eqnarray*}
\langle x,z,\kappa|\opwk{f}|x',z,\kappa\rangle &=& \sum_{m\in\Z^2}
(-)^{Nm_1 m_2} e^{i\pi N\langle m,x'\rangle}
e^{-i(m_1\kappa_1-m_2\kappa_2)}\cr
&{}&~~~~~~~~~~~\langle x,z |\opw{f}|x'-m,z\rangle \;.
\end{eqnarray*}
(ii) For all $n$ in $\Z^2$, for all $x\in\tor$, we have that
$$
\langle x, z,\kappa|U_\kappa(\frac{n}{N})|x, z, \kappa\rangle =
\sum_{m\in\Z^2}  c^m_n(x) e^{-i(m_1\kappa_1-m_2\kappa_2)}\;,
$$
where
$$
c^m_n(x) =  \chi_{n-Nm}(x)\, (-)^{Nm_1 m_2} e^{-i\pi<m,n>} e^{-\pi
N (m-\frac{n}{N},B(z)(m-\frac{n}{N}))} \;.
$$

\noindent (iii) There exists $C>0$ such that
for each $n\in\Z^2$ and for each $x\in \tor$ the following
inequality holds
$$
\left| \langle x, z,\kappa|U_\kappa(\frac{n}{N})|x, z,
\kappa\rangle - c_{n}^{p(\frac{n}{N})}
e^{-i(p\left(\frac{n_1}{N})\kappa_1-p(\frac{n_2}{N})\kappa_2\right)}
\right| \leq C\exp -\frac{\pi N\beta_-(z)}{4}. $$
\end{proposition}
\smallskip
\noindent\dim Part (i) is easily obtained by calculating the Fourier
coefficients of the function $\langle
x,z,\kappa|\opwk{f}|x',z,\kappa\rangle$ in $\kappa$. Indeed, for
$m\in\Z^2$ let
$$
c^{m}(f) = \int_{T^2_\kappa} \frac{d\kappa}{(2\pi)^2} \langle
x,z,\kappa |\opwk{f}| x',z, \kappa\rangle
\e{i(m_1\kappa_1-m_2\kappa_2)}\;.
$$
Then, by using the direct sum decomposition of $L^2(\R)$, of the coherent
states
and of $\opw{f}$ we have
\begin{eqnarray*}
c^{m}(f)
&=& \int_{T^2_\kappa} \frac{d\kappa}{(2\pi)^2}
\langle x,z, \kappa |\opwk{f}U(-m_1,0)U(0,-m_2)|x',z,
\kappa\rangle \\
&=& (-)^{Nm_1 m_2} e^{i\pi N<m,x'>} \int_{T^2_\kappa}
\frac{d\kappa}{(2\pi)^2} \langle x,z, \kappa
|\opwk{f}|x'-m,z,\kappa\rangle \\
&=& (-)^{Nm_1 m_2} e^{iN\pi <m,x'>}
\langle x,z|\opw{f}|x'-m,z\rangle \;.
\end{eqnarray*}
(ii) This is obtained from (i) using the explicit formula of the scalar
product
between two coherent states on the plane (\ref{scalar_cs}).

\noindent(iii) According to the definition given in
(\ref{scalar_cs}) we have that
\begin{eqnarray*}
|\langle x,z,\kappa|U_\kappa(\frac{n}{N})|x,z,\kappa\rangle -
c_n^{p(\frac{n}{N})}
e^{-i(p\left(\frac{n_1}{N})\kappa_1-p(\frac{n_2}{N})\kappa_2\right)}|
&\leq& \sum_{m\neq p(\frac{n}{N})} e^{-\pi N
(m-\frac{n}{N},B(z)(m-\frac{n}{N}))}\cr
&\leq& \sum_{m\neq p(\frac{n}{N})}
e^{-\pi N \beta_-(z)\norm{m-\frac{n}{N}}^2}   \;.
\end{eqnarray*}
The result then comes from the following inequality: let $a>0$ and
$x_o = p(x_o)+f_o\in\R$, with $0\leq|f_o|\leq 1/2$; then we have
\begin{eqnarray*}
\sum_{n\neq p(x_o)}e^{-a(n-x_o)^2}&=&\sum_{n\neq 0}e^{-a(n-f_o)^2}
\leq  \sum_{n\neq 0} e^{-\frac{a}{2}|n-f_o|}\\
&=& \frac{e^{-\frac{a}{2}}}{1-e^{-\frac{a}{2}}}\,
(e^{-\frac{a}{2}f_o}+e^{\frac{a}{2}f_o}) \leq \frac{2 \,
e^{-\frac{a}{4}}}{1-e^{-\frac{a}{2}}} \, . \fin
\end{eqnarray*}
Notice in the above proof that it is convenient to estimate
various $\kappa$-dependent quantities -- which is an a priori
difficult task -- by estimating the terms of the corresponding
$\kappa$-Fourier series expansion, written in terms of quantities
defined on the plane which are often easy to calculate. This
technique will be extensively used in the rest of the paper.

The following estimate makes precise the statement given in the
introduction about quantum fluctuations of the Husimi distribution of
a state $\psi$, both on the plane and on the torus.

\begin{proposition}
\label{husimi} Let $z\in \C, \hbox{ Im }z>0$.

{\rm($i$)}{\rm [Plane]} For each $\psi\in L^2(\R)$,
we have that
\begin{eqnarray*}
\int_{\R^2} |\partial_q|\langle\psi |x,z\rangle|^2|
\frac{dx}{2\pi\hbar} &\leq& |z| \sqrt{\frac{2}{\hbar\ {\rm Im}z}}
\,\norm{\psi}^2  \cr \int_{\R^2} |\partial_p|\langle\psi |
x,z\rangle|^2| \frac{dx}{2\pi\hbar}
&\leq& \sqrt{\frac{2}{\hbar\ {\rm Im}z}} \,\norm{\psi}^2
\end{eqnarray*}

{\rm($ii$)}{\rm [Torus]} For each $\kappa_0\in
[0,2\pi)\times[0,2\pi)$, $\psi_{\kappa_0}\in{\cal H}_N(\kappa_0)$
we have that
\begin{eqnarray*}
\int_{\tor} |\partial_q|\langle\psi_{\kappa_0}
|x,z,\kappa_0\rangle|^2| \frac{dx}{2\pi\hbar} &\leq& 2|z|
\sqrt{\frac{\pi N}{ {\rm Im}z}} \,\norm{\psi_{\kappa_0}}^2_{{\cal
H}_N} \cr \int_{\tor} |\partial_p|\langle\psi_{\kappa_0}
|x,z,\kappa_0\rangle|^2| \frac{dx}{2\pi\hbar} &\leq& 2
\sqrt{\frac{\pi N}{ {\rm Im}z}} \,\norm{\psi_{\kappa_0}}^2_{{\cal
H}_N}
\end{eqnarray*}

\end{proposition}

{\it Proof.} (i) By direct computation we obtain that ($x=(q,p)$)
$$
\partial_q \eta_{x,z} = -\frac{i}{2\hbar} p \eta_{x,z}- \frac{iz}{\hbar} U(x)
Q\ \eta_{0,z}\;, ~~~
\partial_p \eta_{z,x} = \frac{i}{2\hbar} q \eta_{z,x}+ \frac{i}{\hbar} U(x)Q\
\eta_{z,0}\;,
$$
so that
\begin{eqnarray}
\label{deriv_cs}
\partial_q |\langle \psi|x,z\rangle|^2 &=& \frac{2}{\hbar}\ {\rm
Im}\left[ z \langle x,z|\psi\rangle
\langle\psi|U(x)Q|0,z\rangle\right] \cr
\partial_p |\langle \psi|x,z\rangle|^2 &=& -\frac{2}{\hbar}\ {\rm
Im}\left[ \langle x,z|\psi\rangle \langle\psi|U(x)Q|0,z\rangle
\right] \;.
\end{eqnarray}
Then we have
$$
\int_{\R^2}|\partial_q |\langle \psi|x,z\rangle|^2|
\frac{dx}{2\pi\hbar}\qquad\qquad\qquad\qquad\qquad\qquad\qquad
$$
\begin{eqnarray*}
\qquad\qquad\qquad&\leq& \frac{2}{\hbar} \ |z| \int_{\R^2}|\langle
\psi|x,z\rangle \langle \psi|U(x)Q|0,z  \rangle |
\frac{dx}{2\pi\hbar} \cr {} \qquad\qquad\qquad&\leq&
\frac{2}{\hbar} \ |z| \left[\int_{\R^2}|\langle
\psi|x,z\rangle|^2\frac{dx}{2\pi\hbar}\right]^{1/2}
\left[\int_{\R^2}|\langle \psi|U(x)Q|0,z \rangle |^2
\frac{dx}{2\pi\hbar}\right]^{1/2} \cr {} \qquad\qquad\qquad &=&
\frac{2}{\hbar} |z| \norm{\psi}^2 \norm{Q\eta_{0,z}}  \;,
\end{eqnarray*}
where we used (\ref{gen_res_id}). The result then follows from
$\norm{Q\eta_{0,z}}=\sqrt{\frac{\hbar}{2{\rm Im}z}}$.

(ii) By using the Fourier transform in $\kappa$ it is easy to show
that the analogues of equations (\ref{deriv_cs}) are still valid
in ${\cal H}_\hbar(\kappa_0)$, {\it i.e.}
\begin{eqnarray*}
\partial_q |\langle \psi_{\kappa_0}|x,z,\kappa_0\rangle|^2 &=& 4\pi N\ {\rm
Im}\left[ z \langle x,z,\kappa_0|\psi_{\kappa_0}\rangle
\langle\psi_{\kappa_0}|[U(x)Q\eta_{0,z}](\kappa_0)\rangle\right]
\cr
\partial_p |\langle \psi_{\kappa_0}|x,z,\kappa_0\rangle|^2 &=& -4\pi N\ {\rm
Im}\left[ \langle x,z,\kappa_0|\psi_{\kappa_0}\rangle
\langle\psi_{\kappa_0}|[U(x)Q\eta_{0,z}](\kappa_0)\rangle \right]
\;,
\end{eqnarray*}
where $[U(x)Q\eta_{z,0}](\kappa) \cong S(\kappa)U(x)Q\eta_{z,0}$.
The result is then obtained exactly as in (i), by the use of Lemma
\ref{gen_res_id_tor}. \fidi


\bigskip
\bigskip
\sect{The quantized toral automorphisms}
Recalling the quantization
of quadratic hamiltonians on $\R^2$ it is easy to be convinced
that to any linear map specified by $A\in\slr$ Weyl quantization
associates a unitary propagator in $L^2(\R)$
\begin{equation}
(M(A)\psi)(y) = \left(\fraz{i}{2\pi\hbar a_{12}}\right)^{1/2}\int
d\eta\ \e{\frac{i}{\hbar} S_A(y,\eta)} \psi(\eta)  \;,
\label{metapl}
\end{equation}
where $S_A$ is the classical action associated to $A$ (see \cite{f} for
details). A
fundamental property of Weyl quantization with respect to this
dynamics is the absence of an error term in the so called {\it
Egorov theorem}, meaning
 that ``quantization and  evolution commute'', i.e.
\begin{equation}
\label{egorov} \ma{-t} \opw{f} \ma{t} = \opw{f\circ A^t}\ \forall
t\in\Z \;.
\end{equation}

The quantization of $A\in\slz$ on the torus is straightforward. It
turns out that for all $N$ there exists $\kappa$ such that
$M(A)\hk\subset\hk$ (see \cite{bodb2} for the equation that
$\kappa$ must satisfy). We shall write $M_\kappa(A)$ for the
restriction of $M(A)$ to $\hk$. From this construction it follows
that (\ref{egorov}) holds for $Op^W_\kappa(f)$ as well. We finally
recall  that $M(A)\mid x,z \rangle = \e{i\phi(A)}\mid Ax, A\cdot z
\rangle$ where $A\cdot z= (za_{22}+a_{21})/(za_{12}+a_{11})$
(closely related to the usual homographic action),  and where $\phi(A)$
is a phase we don't specify. As a result,
\begin{equation}
M_\kappa(A)\mid x,z,\kappa \rangle = \e{i\phi(A)}\mid Ax, A\cdot
z, \kappa \rangle.
\end{equation}

\subsection{The proof of Theorem \ref{cat}}
\bigskip
Our strategy for the first proof of (\ref{classical_limit_intro})
and (\ref{mixing_limit_intro}) will be as follows. We will first
consider
\begin{equation}
O_f^{W,\kappa}[t,N] (x_0)= \langle x_0,z,\kappa |
\mak{-t}\opwk{f}\mak{t} |x_0,z, \kappa\rangle
\end{equation}
rather than $O_f^{AW,\kappa}[t,N](x_0)$ (see (\ref{observable_t}))
and prove the results for those objects (Propositions
\ref{plane_limit}-\ref{torus_limit}).
 Equations
(\ref{classical_limit_intro}) and (\ref{mixing_limit_intro}) then follow
immediately from the  observation
\cite{bodb2} that
$$
||Op^{AW}_{\kappa,z}(f) - Op^W_\kappa(f)||\leq C_f \frac{1}{N}.
$$
To study the limiting behaviour of $O_f^{W,\kappa}[t,N]$, we
proceed in two steps: first, we show  the limiting behaviour of
$O_f^{W,\kappa}[t,N](x_0)$ is identical to that of
\begin{equation}
O_f^{W}[t,N](x_0) = \langle x_0,z | \ma{-t}Op^W{(f)}\ma{t} |x_0,z
\rangle
\end{equation}
for times up to $\frac{\ln N}{\gamma}$ (Proposition
\ref{torus_limit}); here $\ma{}$ is defined in (\ref{metapl}) and
the coherent states $|x_0, z\rangle$ in (\ref{cs}). Hence the
problem is reduced to the analysis of the limiting behaviour of
$O_f^{W}[N,t]$, which is essentially trivial since $M(A)$ is just
the quantization of the ordinary linear dynamics on the full phase
plane, rather than on the torus (Proposition \ref{plane_limit}).

Let $A\in\slr$ with $|{\rm Tr}A|> 2$. We write $Av_{\pm} =
\lambda^{\pm 1} v_{\pm}, \lambda=\exp\gamma>1,
v_{\pm}=(\cos\theta_\pm, \sin\theta_{\pm})$ for the eigenvectors
and the eigenvalues of $A$. We will always write
$N=1/(2\pi\hbar)$, although $N$ will be assumed to be an integer
only when we deal with the quantum map on $\hk$ (in which case
automatically $A\in \slz$), in Proposition \ref{torus_limit}.

\begin{proposition}
\label{plane_limit} Let $z\in\C$, with ${\rm Im}z>0$.

\noindent (i) Let $A\in \slr$ with $|{\rm Tr}A|>2$. Then, for each
$f\in\l{2}$, there
exists
$C_f>0$ such that $\forall\,x_0\in\tor$
$$
\left| O_f^W[t,N](x_0) - (f\circ A^t)(x_0) \right| \leq C_f
\beta_+(z) \fraz{\lambda^{2t}}{N} \;.
$$
(ii) Let $A\in \slz$ with $|{\rm Tr}A|>2$. If $f\in\l{k},\ (k\in\N^*)$ then
there
exist $C_{f,1},C_{f,2}>0$ such  that $\forall\,x_0\in\tor$ and
$\forall\,(t,N)$
$$
\left| O_f^W[t,N](x_0)-\int_{\tor}dx\,\,f(x) \right| \leq
\frac{C_{f,1}}{\beta_-(z)^k}
\left(\fraz{N}{\lambda^{2t}}\right)^{\frac{k}{2}} + C_{f,2}
\,\e{-\pi\fraz{\lambda^{2t}}{N}c^2_+\beta_-(z)}\;,
$$
where $c_+={\rm Max}\{|\cos\theta_+|,|\sin\theta_+|\}$.
\end{proposition}
\noindent\dim (i)
Since $\chi_\xi(x)=\exp 2\pi i\langle x, \xi\rangle$, we have
$\opw{\chi_{\xi}}=\U{\xi_1/N}{\xi_2/N}$ and hence we obtain, using
(\ref{scalar_cs}),
\begin{equation}
\label{plane_waves} O_{\chi_{\xi}}^W[0,N](x_0) = \e{2i\pi\langle
x_0,\xi\rangle} \e{-\frac{\pi}{N}(\xi,B(z)\xi)}\;.
\end{equation}
Using (\ref{egorov}), it then follows that $O_{\chi_\xi}^W[t,N] =
O_{\chi_{\xi_t}}^W[0,N]$ where
$
\xi_t= A^{-t}\xi\;.
$
As a result, if $f=\sum_{n}f_{n} \chi_{n}$, then
\begin{equation}
\label{form_gen} O_f^W[t,N](x_0) = \sum_{n\in\Z^2} f_{n}
\e{-\frac{\pi}{N} (n_t,B(z)n_t)}\chi_{n_t}(x_0)\;.
\end{equation}
Using (\ref{form_gen}) we have that
\begin{eqnarray*}
\left| O_f^W[t,N](x_0) - f\circ A^t(x_0) \right| &\leq&
\sum_{n\in\Z^2} |f_{n}| (1-\e{-\frac{\pi}{N} (n_t,B(z)n_t)}) \cr
&\leq &  \frac{\pi}{N}\sum_{n\in\Z^2}|f_{n}|(n_t,B(z)n_t)\;, \cr
&\leq & \frac{\pi\beta_+(z)}{N} \sum_{n\in\Z^2} |f_n|\norm{n_t}^2
\;,
 \end{eqnarray*}
where we used the inequality $1-\e{-y}\leq y$, for each $y\geq 0$.
The result then follows from the properties of $\|A^{-t}\|$ and
from the regularity of $f$.

\noindent (ii) We
recall that $s_+=\cot\theta_+$ is a quadratic irrational. The
basic properties of quadratic irrationals that we will use can be
found in \cite{kh}. Using the Schwarz inequality we have that
$$
\left\|A^{-t} {n}\right\|\geq
\left|\langle v_+,A^{-t}{n}\rangle\right|=
\left|\langle A^{t}v_+,{n}\rangle\right|=
\lambda^{t}|\sin\theta_+||n_1-n_2 s_+| \;,
$$
and from (\ref{form_gen}) we obtain the following estimate
$$
\left| O_f^W[t,N](x_0)-\int_{\tor}dx\,f(x) \right| \leq
\sum_{{n}\neq{\bf 0}} |f_{n}| \,\e{-\pi \beta_-(z)
\fraz{\lambda^{2t}}{N} \sin^2\theta_+|n_1-n_2 s_+|^2}\;.
$$
Without loss of generality suppose that $s_+>0$. It is convenient
to divide the sum in  two parts, $I_1$ and $I_2$. In $I_1$ we sum
over $n_1\cdot n_2\leq 0$, from which we easily obtain the
exponential term in the estimate. To discuss $I_2$, we recall that
since $s_+$ is a quadratic irrational, there exists a constant
$C>0$ such that $\forall \,n_1,n_2>0$
$$
\left|n_1-s_+n_2\right| > \fraz{1}{C n_2}  \;.
$$
Then,
$$
I_2 \leq \sum_{n_1,n_2>0} \left(|f_{n_1,n_2}|+|f_{-n_1,-n_2}|
\right) \e{-\pi\beta_-(z)\fraz{\lambda^{2t}}{N}\fraz{\sin^2
\theta_+}{C^2 n_2^2}}\,.
$$
We then obtain the final estimate using the inequality $\exp(-x) < C'_k/x^k$,
valid for some $C'_k>0$ and for each $x>0$. \fin

\ From now on, $N=1/(2\pi\hbar)$ is an integer and let $\hh$ be the Hilbert
space defined in section 2.
The key observation allowing a simple analysis of the behaviour of
$O^{W, \kappa}_f[t,N]$ in the $(t,N)$-plane is the following proposition,
which
shows that for times $t<<\frac{\ln N}{\gamma}$, $O^{W, \kappa}_f[t,N]$ is
close
to $O_f^W[t,N]$, so that its behaviour can be read off from
Proposition \ref{plane_limit} above.

\smallskip
\begin{proposition}
\label{torus_limit} Let $z\in\C$, with ${\rm Im}z>0$. If
$f\in\l{k}$, then there exist $C_{f,1},C_{f,2}>0$ such that for
each $x_0\in\tor$ and all $t,N$
$$
\left| O^{W, \kappa}_f[t,N](x_0)-O^W_f[t,N](x_0) \right| < C_{f,2}
\left(\fraz{\lambda^t}{N}\right)^k + C_{f,1}\exp -\frac{\pi
N}{4}\beta_-(z) \;.
$$
\end{proposition}

\noindent\dim
Using the inequality of Proposition \ref{fourier_gen}(iii) and
recalling that $c_{n}^{0}=O_{\chi_n}^W[0,N]$, it is easy to find
$C_{f,1}$ such that
$$
\left| O^{W,\kappa}_f[t,N](x_0)-O^{W}_f[t,N](x_0) \right| \leq
C_{f,1} e^{-\frac{\pi N}{4}\beta_-(z)}+
$$
$$
\qquad\qquad\qquad\qquad\qquad\qquad\sum_{n}|f_{n}| \left|
c_{n_t}^{p(\frac{n_t}{N})}(x_0)
e^{-i(p\left(\frac{{n_t}_1}{N})\kappa_1-p(\frac{{n_t}_2}{N})\kappa_2\right)}
- c^{0}_{n_t}(x_0) \right|\;.
$$
Because $p(x)\neq 0$ only if $|x|>1/2$ the sum is limited to ${n}$ such
that $|{n_t}_1/N|$ or $|{n_t}_2/N| >1/2$. Then, we have
\begin{eqnarray*}
\left| O^{W, \kappa}_f[t,N](x_0)-O_f^W[t,N](x_0) \right| &\leq &
C_{f,1} e^{-\frac{\pi N}{4}\beta_-(z)} + 2
\sum_{\|A^{-t}n\|>\frac{N}{2}} |f_{n}| \\
&\leq& C_{f,1} e^{-\frac{\pi N}{4}\beta_-(z)} + 2 \sum_{n} |f_{
n}|\left(\frac{2\norm{A^{-t}n}}{N}\right)^k   \\
&\leq & C_{f,1}e^{-\frac{\pi N}{4}\beta_-(z)} +
2 \left(\fraz{2\|A^{-t}\|}{N}\right)^k \sum_{ n}|f_{ n}| \|{ n}\|^k \;.
\fidi\end{eqnarray*}

\noindent It is now clear that the results of Propositions \ref{plane_limit}
and \ref{torus_limit} imply Theorem \ref{cat}.


We now give a more direct proof of the mixing regime of
Theorem \ref{cat} for the Husimi distribution, based on the intuitive picture
presented in the introduction and in particular on (\ref{expmix}).

\begin{proposition}
\label{direct_proof} Let $z\in\C$, {\rm (${\rm Im}z>0$)}, and
$A\in{\rm SL_2(\Z)}$, with ${\rm Tr}A>2$.

{\rm ($i$)} Let $f\in \l{1}$. There exists $C_{f,A}>0$ such that
$\forall x_0\in\R^2$, $\hbar,t>0$,
$$
|\langle x_0,z|M(A)^{-t} \opaw{f}M(A)^t|x_0,z\rangle - \int_\tor
dx \ f(x)| \leq C_{f,A} \frac{|z|+1}{\sqrt{{\rm
Im}z}}\frac{1}{\lambda^t \sqrt{\hbar}} \;.
$$

{\rm ($ii$)} Let $f\in \l{k}$, with $k\geq 1$. There exist
$C_{f,A}, C_{f_1}, C_{f_2}>0$ such that $\forall x_0\in \tor$,
$N,t>0$
$$
|\langle x_0,z,\kappa|M_\kappa(A)^{-t}
\opawk{f}M_\kappa(A)^t|x_0,z,\kappa\rangle - \int_\tor dx \ f(x)|
$$
$$
\qquad\qquad\qquad\qquad\qquad\qquad\qquad\leq C_{f,A}
\frac{\sqrt{N}}{\lambda^t} + C_{f_2}
\left(\frac{\lambda^t}{N}\right)^k + C_{f_1} e^{-\pi
N\beta_-(z)/4} \;.
$$
\end{proposition}

{\it Proof.}($i$) Let $f_1=\sum_n f_{1,n}\chi_n\in \l{1}$ (with
$\int_\tor dx\ f_1(x)=0$) and $f_2\in C^\infty(\R^2)$
such that $\int_{\R^2}dy\ [|\partial_q f_2|+|\partial_p f_2
|]<\infty$. Adapting the proof given in Theorem 4 of \cite{db} it
can be shown that
$$
|\int_{\R^2} dy\ (f_1\circ A^t)(y) f_2(y)| \leq C_A \lambda^{-t}
(\sum_n |f_{1,n}| \norm{n}) \int_{\R^2}dy\ [|\partial_q
f_2|+|\partial_p f_2 |] \;,
$$
for some $C_A>0$. Then, if $\int_\tor f =0$
\begin{eqnarray*}
|\langle x_0,z|M(A)^{-t} \opaw{f}M(A)^t|x_0,z\rangle| &=&
|\int_{\R^2} dy\ (f\circ A^t)(y) \frac{|\langle y, A^{-t}\cdot
z|x_0,z\rangle|^2}{2\pi\hbar}| \cr
&=&|\int_{\R^2} dy\ (f\circ A^t)(y) \frac{|\langle 0, A^{-t}\cdot
z|x_0-y,z\rangle|^2}{2\pi\hbar}| \cr
&=& |\int_{\R^2} d\xi\ (f\circ A^t)(x_0-\xi)
\frac{|\langle 0, A^{-t} \cdot z|\xi,z\rangle|^2}{2\pi\hbar}| \cr
&\leq& C_A (\sum_n |f_n| \norm{n}) \lambda^{-t}
\int_{\R^2}dy \ [|\partial_q h|+|\partial_p h|]  ,
\end{eqnarray*}
where $h(\xi)=|\langle 0, A^{-t} \cdot
z|\xi,z\rangle|^2/(2\pi\hbar)$ is the Husimi distribution of
$|\psi\rangle = |0,A^{-t}z\rangle$. The result then follows from
Proposition \ref{husimi}($i$).

($ii$) The only difficulty to repeat on the torus the estimate
given under ($i$) comes from the fact that $|\langle
y,z_1,\kappa|x,z_2,\kappa\rangle|^2$ $\neq |\langle
0,z_1,\kappa|x-y,z_2,\kappa\rangle|^2$. Indeed,
$$
\langle x_0,z,\kappa|M(A)^{-t} \opawk{f}M(A)^t|x_0,z,\kappa\rangle
= ~~~~~~~~~~~~~~~~~~~~~~~
$$
$$
\int_{\tor} dy\ (f\circ A^t)(x_0-y) \frac{|\langle 0, A^{-t}\cdot
z;\kappa|y,z;\kappa\rangle|^2}{2\pi\hbar}  + \int_\tor dy \
G(x_0,z;y,A^{-t}\cdot z) (f\circ A^t)(y) \;,
$$
where $G(x,z_1;y,z_2)=N (|\langle
y,z_2,\kappa|x,z_1,\kappa\rangle|^2 - |\langle
0,z_2,\kappa|x-y,z_1,\kappa\rangle|^2)$. The first term can be
evaluated as in part ($i$), using this time Proposition
\ref{husimi} ($ii$). Let $G(x,z_1;y,z_2)=\sum_\beta
G_\beta(x,z_1;z_2)\exp i2\pi\langle\beta,y\rangle$. By using the
definition of anti-Wick quantization we easily rewrite
$$
G_\beta(x,z_1;z_2) = \langle
x,z_1,\kappa|Op^{AW}_{\kappa,z_2}(\chi_\beta)|x,z_1,\kappa\rangle
-\chi_\beta(x) \langle
0,z_2,\kappa|Op^{AW}_{\kappa,z_1}(\chi_\beta^*)|0,z_2,\kappa
\rangle \;.
$$
By subtracting the same quantity calculated on the plane, which is
zero, and by using Lemma \ref{weyl_antiW} we finally have
\begin{eqnarray*}
&&G_\beta(x,z_1;z_2) = \langle \frac{\beta}{N},z_2|0,z_2\rangle
\left[\langle x,z_1,\kappa|\opwk{\chi_\beta}|x,z_1,\kappa\rangle -
\langle x,z_1|\opw{\chi_\beta}|x,z_1\rangle \right] \cr
&&~~~~- \chi_\beta(x) \langle 0,z_1|\frac{\beta}{N},z_1\rangle
\left[\langle 0,z_2,\kappa|\opwk{\chi_\beta^*}|0,z_2,\kappa\rangle
- \langle 0,z_2|\opw{\chi_\beta^*}|0,z_2\rangle \right]\;.
\end{eqnarray*}
After some simple calculation we can finally write
($\beta_t=A^{-t}\beta $)
\begin{eqnarray*}
& &\int_\tor dy \ G(x_0,z;y,A^{-t}\cdot z) (f\circ A^t)(y) =\cr
& &~~~~\sum_\beta f_\beta \langle \frac{\beta}{N},z|0,z\rangle
\left[\langle x_0,z,\kappa|\opwk{\chi_{\beta_t}}|x_0,z,\kappa\rangle -
\langle x_0,z|\opw{\chi_{\beta_t}}|x_0,z\rangle \right]-\cr
& & ~~\sum_\beta f_\beta\chi_{\beta_t}(x_0)\langle
0,z|\frac{\beta_t}{N},z\rangle \left[\langle
0,z,\kappa|\opwk{\chi_{\beta}^*}|0,z,\kappa\rangle - \langle
0,z|\opw{\chi_{\beta}^*}|0,z\rangle \right]\;.
\end{eqnarray*}
The result then follows by making use of Proposition
\ref{torus_limit}. \fidi



\subsection{The Period of the Quantum Map}\label{sec_period}
It is natural to wonder if (\ref{mixing_limit_intro}) holds beyond
$\frac{\ln N} {\gamma}$ and in particular if it holds possibly for
times that are polynomial in $N$. We shall show (see
(\ref{mixlong})-(\ref{instablong})) with an example that the
mixing regime may break down on a time scale which is only
logarithmic in $N$, thereby showing that on this scale the
agreement between the classical and quantum or semi-classical
evolutions breaks down.

The intuitive ``proof" of (\ref{mixing_limit_intro}) given in the
introduction, which is based on (\ref{expmix}), does not predict
any breakdown of (\ref{mixing_limit_intro}) at long times.  It is
possible to get an intuitive understanding of the breakdown of the
mixing regime using the uncertainty principle as follows.  First
remark that, given any $\psi\in\hk$, the support of its Wigner
distribution has, due to the uncertainty principle, necessarily a
linear size of the order of at least $\hbar$ in all directions:
indeed, from $\Delta X \Delta P \geq \hbar$ and $\Delta X, \Delta
P \leq 1$, one concludes $\Delta X, \Delta P \geq \hbar$. Consider
now $\ma{t}|x, z, \kappa\rangle$; we expect its Wigner
distribution to have a spread of $\sqrt \hbar \exp \gamma t$ along
the unstable direction and therefore to wrap around the torus as
soon as $\sqrt \hbar \exp \gamma t>> 1$. The transversal distance
between the successive windings so obtained can be estimated by
$\sqrt \hbar^{-1} \exp -\gamma t$. When this distance becomes less
than $\hbar$, the support of the Wigner distribution of $\ma{t}|x,
z, \kappa\rangle$ can no longer separate the separate windings,
because of the previous remark, and as a result one expects the
classical evolution picture may break down at such times, given by
$\sqrt \hbar^{-1} \exp -\gamma t \sim \hbar$ or $t\sim
\frac{3}{2}\frac{|\ln\hbar|}{\gamma}$. We shall now exhibit
precisely such a phenomenon for the quantized cat maps.

For that purpose we consider $A$ such that either $a_{12}=1$ or
$a_{21}=1$ and $[a_{11}]_2=[a_{22}]_2=0$ (we use the notation
$[x]_n=x \;{\rm mod}\,n$, both for numbers and matrices). We know
that $\kappa=0$ gives an admissible quantization; furthermore it
was shown in \cite{hb} that the corresponding quantum map $M_o(A)$
is periodic with period $n(N)={\rm Min}\,\left\{ t \;|\; M_o(A)^t=
e^{i\phi_N}\,,\;\phi_N\in\R ~ \right\}$. These periods have been
studied in \cite {ke}, were it is argued that they behave ``on
average" linearly in $N$, but with great fluctuations about this
average. It is of course clear that the ``mixing regime'' must
break down before the period.  In the following we will show that
there exists a sequence $N_{2k+1}$ of values of $N$ for which the
period is extremely short: $n(N_{2k+1}) \sim 2\frac{\ln
N_{2k+1}}{\gamma}$, leading to the announced result.

We will need the following simple formulas for $A$.
If we call
$\lambda$  the biggest eigenvalue of $A$, we know that for each $t\in\N^+$
\begin{equation}
\label{iterate} A^t = p_t A - p_{t-1} \;,~~~~p_{t+1} = {\rm Tr}(A)
p_t - p_{t-1}   \;, \hbox{where}\; p_t =
\fraz{\lambda^t-\lambda^{-t}}{\lambda-\lambda^{-1}} \;.
\end{equation}
We also introduce
$T_N = {\rm Min}\,\left\{ t \;|\; [A^t]_N = 1\,\right\}.$
We denote by $A_N$ the matrix
with integer entries such that $A^{T_N}=1+NA_N$.
Then, following \cite{hb}, $n(N)=T_N$ if $N$ is odd or if $N$ is even and
$[(A_N)_{12}]_2=[(A_N)_{21}]_2=0$; otherwise $n(N)=2T_N$.
We finally define for $k\in\N^+$
\begin{equation}
\label{p_series}
N_k \equiv {\rm Max}\,\left\{ N \;|\; [A^k]_N = 1 ~ \right\} \;,
\end{equation}
and we prove the following result.
\begin{proposition}
\label{pro_period}
For each $k\in\N^+$ we have that $T_{N_k}=k$ and
\begin{equation}
\label{period}
N_{2k} = 2 p_k,  ~~~~~~ N_{2k+1} = p_k + p_{k+1} \;.
\end{equation}
\end{proposition}

\noindent\dim Using (\ref{iterate}), we see that $N_k$ is the greatest
integer
such that
\begin{eqnarray*}
&[p_k a_{11} - p_{k-1} - 1]_{N_k}=0 ~~~~~ &
[p_k a_{12}]_{N_k}=0 \cr
&[p_k a_{22} - p_{k-1} - 1]_{N_k}=0 ~~~~~ &
[p_k a_{21}]_{N_k}=0\;,
\end{eqnarray*}
or, because of the hypothesis about the off-diagonal terms of $A$,
$
[p_k]_{N_k} = 0
$
,
$
 [p_{k-1}]_{N_k} = -1.
$
This means that $N_k$ is the greatest common divisor of $p_k$ and
$p_{k-1}+1$, i.e.
$N_k = (p_k, p_{k-1}+1)$. We are going to show by induction that, for each
$s=0,\dots,k-1$,
$$
N_k = (p_{k-s}+p_s,p_{k-(s+1)} + p_{s+1}) \;.
$$
Since $p_0=0$ and $p_1=1$, this is clearly true for $s=0$. Supposing it is
true for
$s$, we have
\begin{eqnarray*}
N_k & = &({\rm Tr}A \,p_{k-s-1}-p_{k-s-2}+p_s, \, p_{k-s-1}+p_{s+1}) \cr
    & = &({\rm Tr}A\,(p_{k-s-1}+p_{s+1})-{\rm Tr}A\,p_{s+1}-p_{k-s-2}+p_s,
\, p_{k-s-1}+p_{s+1}) \cr
    & = &(p_{s+2}+p_{k-(s+2)}, \, p_{k-(s+1)}+p_{s+1}) \;,
\end{eqnarray*}
so that it is true for $s+1$. In the third line we used the identity
$(a,\, ca-b) = (a,\, b)$,
valid for all $a,b,c$, and formula ({\ref{iterate}}).

If $k=2\ell$, then, setting $s=\ell$ in the above formula, we have
$$
N_{2\ell} = (2p_\ell,\, p_{\ell-1}+p_{\ell+1}) = (2p_\ell,\, {\rm Tr}A\,
p_\ell)
 = 2 p_\ell;,
$$
because ${\rm Tr}A$ is pair by hypothesis. If $k=2\ell+1$, then
$
N_{2\ell+1} = p_\ell + p_{\ell+1}  \;.
$
For each $k$ we have that $[A^k]_{N_k}=[A^{T_{N_k}}]_{N_k} =
[A^{T_{N_k}}]_{N_{T_{N_k}}}=1$.
\ From the definition of the period  we have that $T_{N_k} \leq k$ and from
the
definition
of $N_k$ it follows that $N_k\leq N_{T_{N_k}}$. Since the sequence $\{N_k\}$
is
increasing (see (\ref{iterate})-(\ref{period})) we conclude that $T_{N_k}\geq
k$ and hence that $T_{N_k}=k$.
\fidi

\medskip

Since $[{\rm Tr} A]_2=0$, we have that $[p_{2k}]_2=0$ and
$[p_{2k+1}]_2=1$, so that $[N_{2k+1}]_2=1$. Using the results of
Proposition \ref{pro_period}, it then follows that $2k+1$ is the
quantum period for $N_{2k+1}=p_k+p_{k+1}$, i.e.
$n(N_{2k+1})\approx 2\frac{\ln N_{2k+1}}{\gamma}$. Keeping in mind
that  $M(A)^{-1} = M(A^{-1})$, so that $M(A)^t=
M(A)^{t-n(N_{2k+1})}=M(A^{-1})^{n(N_{2k+1})-t)}, \ (0\leq t\leq n(N_{2k+1}))$,
we can apply (\ref{classical_limit_intro}) and
(\ref{mixing_limit_intro}) to $A^{-1}$ to conclude that if we perform the
limits running only over $N_{2k+1}$, we have
\begin{eqnarray}
& &\lim_{\stackrel{k\rightarrow\infty}{ \frac{\ln N_{2k+1}}{2\gamma}\ll t \ll
\frac{3}{2}\frac{\ln N_{2k+1}}{\gamma}}}
O_f^{AW,0}[t,N_{2k+1}](x_0) = \int_\tor f(x') \,dx'  \label{mixlong}\\
& &\lim_{\stackrel{k\rightarrow\infty}{ \frac{3}{2}\frac{\ln
N_{2k+1}}{\gamma}\ll
t \ll 2\frac{\ln N_{2k+1}}{\gamma}}}
|O_f^{AW,0}[t,N_{2k+1}](x_0)-\left(f\circ A^{t-n(N_{2k+1})}\right)(x_0)| = 0
\;.
\label{instablong}
\end{eqnarray}
Equation (\ref{instablong}) clearly shows the breakdown of the mixing regime
for times beyond $\frac{3}{2}\frac{\ln N}{\gamma}$, in the cases considered
here. It is of course still possible that, ``generically", it remains valid
for much longer times, as the classical intuition would predict.


\bigskip
\bigskip
\subsection{Position eigenstates}\label{sec_catpos}
In this subsection we study the Wigner and Husimi distributions of evolved
position
eigenstates by studying the matrix elements
$\langle e^\kappa_{j}, \ma{-t}\opwk{f}\ma{t} e^\kappa_j \rangle$ and
$\langle e^\kappa_{j}, \ma{-t}\opawk{f}\ma{t} e^\kappa_j \rangle$ in the
limits
$t\to \infty$ and $N\to \infty$. If we applied the heuristic argument of the
introduction
to this case, we would conclude that the mixing regime should set in {\em no
later
than} at times of order $\frac{\ln N}{2\gamma}$. As a matter of fact, it sets
in much sooner, as Proposition \ref{pro_catpos} shows. We need the following
preparatory result.
\begin{lemma}
\label{pos_eig}
For each $f\in\l{k}$, there exist $C_{f}>0$ such that for each
$N,t\geq 0$,
$j=0,\ldots N-1$ and $|J|<N/2$ we have that
\begin{eqnarray}
|\langle e^\kappa_{j+J}, \ma{-t}\opwk{f}\ma{t} e^\kappa_j \rangle &-&
\int_0^1 dp\,(f\circ A^t)(\frac{\kappa_2/\pi + 2j + J}{2N},p) e^{i2\pi
Jp}|
\nonumber \\ {} &\leq& C_{f}\left( \frac{\norm{A^{-t}}}{N}\right)^k
\;.
\label{pos_eig_ineq}
\end{eqnarray}
\end{lemma}
\noindent\dim Using the Egorov theorem and (\ref{heisenberg}) we find that
\begin{eqnarray*}
\langle e^\kappa_{j+J},\ma{-t}\opwk{f}\ma{t}e^\kappa_j \rangle &=&
\sum_{n} f_{A^t n} e^{i\pi \frac{n_1n_2}{N}}
e^{i(\kappa_2+2\pi j)\frac{n_2}{N}} \langle
e^\kappa_{j+J},e^\kappa_{j+n_1}\rangle \\
{}&=&
\sum_{n} f_{A^t(J+n_1N,n_2)} e^{-ik_1n_1} e^{i\frac{\pi}{N} n_2(J+Nn_1)}
e^{i(\kappa_2+2\pi j)\frac{n_2}{N}} \;.
\end{eqnarray*}
It is clear that the integral in (\ref{pos_eig_ineq}) is obtained
posing $n_1=0$. The remaining terms are easily estimated:
\begin{eqnarray*}
\sum_{\stackrel{n}{n_1\neq 0}}|f_{A^t(J+Nn_1,n_2)}| &\leq&
\sum_{\norm{n}\geq \frac{N}{2}} |f_{A^tn}| \cr
&\leq& \sum_n
\left(\frac{2\norm{n}}{N}\right)^k |f_{A^tn}|
\leq \left(\frac{2\norm{A^{-t}}}{N}\right)^k \sum_n\norm{n}^k|f_n|
\;.
\fidi
\end{eqnarray*}

\begin{proposition}\label{pro_catpos}
For each $f\in \l{k}$, there exist $C_{f,1},C_{f,2}$ such that,
for all $t,N$, $j=0,\ldots,N-1$, one has
$$
|\langle e^\kappa_{j}, \ma{-t}\opwk{f}\ma{t} e^\kappa_j \rangle
 -
\int_\tor dx\,f | \leq
C_{f,1}\left(\frac{\norm{A^{-t}}}{N}\right)^k +
\frac{C_{f,2}}{\lambda^{tk}} \;.
$$
\end{proposition}

\noindent\dim From (\ref{pos_eig_ineq}) it is clear that it will be enough to
show that it exists $C_{f,2}$ such that, for all $0\leq q \leq 1$,
$$
|\int_0^1(f\circ A^t)(q,p) dp - \int_{\tor} f(x) dx|\leq
\frac{C_{f,2}}{\lambda^{tk}}.
$$
Indeed,
\begin{eqnarray*}
|\int_0^1(f\circ A^t)(q,p) dp - \int_{\tor} f(x) dx|&\leq&
\sum_{n\neq 0}
|f_{A^t(0,n)}|\leq\sum_{\norm{m}\geq\lambda^t|\cos\theta_-|}|f_m|
\cr
& &\leq\frac{1}{\lambda^{tk}|\cos\theta_-|^k}\sum_m|f_m|\norm{m}^k
\;,
\end{eqnarray*}
where we used the inequality $\norm{A^t(0,n)}\geq|\langle
v_-,A^t(0,n)\rangle|\geq\lambda^t|\cos\theta_-|$, valid for each
$n\neq 0$. \fidi

\noindent It is clear that we actually have, for any sequence
$\theta_N\to\infty$,
$$
\lim_{\stackrel{N\to\infty}{\theta_N<t<(1-\epsilon)\frac{\ln
N}{\gamma}}} \langle e^\kappa_{j}, \ma{-t}\opwk{f}\ma{t}
e^\kappa_j \rangle
=
\int_\tor f(x)\ dx   \;.
$$
Therefore there is no ``orbital instability regime" for initial
position eigenstates, but this is easily understood in terms of
the classical dynamics applied to the support of the initial
Wigner distribution.


\sect{The Quantum Baker Map}\label{qbaker}
\subsection{Presentation of the model and of the results}\label{qbaker1}

The Baker map is a discontinuous map $B$ on the torus, defined on
$x=(q,p)\in\tor$
by
\begin{equation}
\label{baker_classic}
B x = \left\{
\begin{array}{cc}
B_o x & q\in[0,1/2[\cr
B_ox + (-1,1/2) & q\in]1/2,1]
\end{array}
\right. \;,
\end{equation}
where $B_o$ is the linearized map given by
$$
\left(
\begin{array}{cc} 2 & 0\cr 0 & \frac{1}{2}\end{array}
\right)\;.
$$
The following relation concerning the action of $B$ on the characters
$\chi_n$,
$n\in\Z^2$, will be crucial:
\begin{equation}
\label{baker_characters}
\chi_{2n_1,n_2}\circ B = \chi_{n_1,2n_2}  \;\;.
\end{equation}

With the usual assumption $[N]_2=0$ and posing $\kappa=0$ (see \cite{bavo} or
\cite{dbde} and further references therein for the general scheme) the
quantization of $B$
is given by the following unitary operator on ${\cal H}_N\equiv{\cal
H}_N(0)$:
$$
V_B = \eff_N^{-1}\circ
\left(\begin{array}{cc}\eff_{N/2} & 0 \cr 0 & \eff_{N/2}\end{array}
\right)
$$
where
$\begin{displaystyle}
(\eff_{N})_{k,\ell} = \frac{1}{\sqrt{N}} \e{-i2\pi\frac{k\ell}{N}}
\end{displaystyle}$
are the matrix elements of $\eff$ in the position eigenstate basis
$\{e_j\equiv e^0_j\}_{j=0}^{N-1}$.

In this section we will mean by semi-classical evolution of the Husimi
function associated with $\psi\in {\cal H}_N$ the following function
\begin{equation}
\label{s_class_husimi} g^{sc}_{\psi}(x,t) = \frac{|\langle
B^{-t}x,B^{-t}_o\cdot z | \psi\rangle|^2}{2\pi\hbar} \;,
\end{equation}
where $B^{-t}_o\cdot z=4^t z$. We will write $g^{sc}_j$, respectively
$g^{sc}_{x_0,z}$ when $\psi=e_j$, respectively $\psi=|x_0,z;0\rangle$. We will
compare this quantity to the Husimi distribution of the quantum mechanically
evolved state.

Since throughout the rest of this section $\kappa=0$, and in order to
alleviate the
notations, we will  drop the $\kappa$-dependence of many symbols.
In the following propositions we list our principal results about the quantum
and semi-classical
evolutions of coherent states and position eigenstates from which Theorem
\ref{baker} will follow easily. The proofs are
postponed to the next subsection. From now on, we will deal with observables
depending only on the $q$-variable, for which we will write $f\in
C^\infty(\T)$.

We consider for each $N$ a lattice of $M_N<N$ points $x^{(N)}_j$
on the torus, given by $x_j^{(N)}=(\frac{j_1}{\sqrt{M_N}},
\frac{j_2}{\sqrt{M_N}}),\ 1\leq j_1, j_2 \leq \sqrt{M_N}$.
Proposition \ref{fourier_gen} (i) with $f=1$ and $z=i$ then
immediately implies
\begin{equation}\label{cohstateoverlap}
|\langle x_k^{(N)}, i|x_j^{(N)}, i\rangle -\delta_{kj}| \leq C\
\exp -\frac{\pi}{2}\frac{N}{M_N},
\end{equation}
where we wrote, with some abuse of notation  $|x_j^{(N)}, z\rangle
\equiv |x_j^{(N)}, z, 0\rangle\in {\cal H}_\hbar(0)\sim {\cal H}_\hbar$ (Do
not confuse with (\ref{cs})). Since we will need to apply Lemma \ref{gram}
below to
the $|x_j^{(N)},i>$, we will always need the sequence $M_N$ to be
such that
\begin{equation}\label{mncondition}
\label{cs_diseq} 8CM_N\exp-\frac{\pi}{2}\frac{N}{M_N}\leq 1.
\end{equation}
Consequently, from now on, we will always take
 $\frac{\pi}{2}\frac{N}{\ln 8CN}-1 \leq M_N \leq \frac{\pi}{2}\frac{N}{\ln
8CN}$.
We can now state the main ingredient for the proof of Theorem \ref{baker}:
\begin{proposition}
\label{q_baker_cs} {\rm [Coherent states]} {\rm ($i$)} Let $N>0,
0< \epsilon < 1$ and $0<\alpha<\frac{\epsilon}{4}$ be given and
let $M_N$ be as above. Then there exists a subset ${\cal M}_N$ of
the index set $1\leq i_1, i_2\leq \sqrt{M_N}$ with the following
properties:
\begin{itemize}
\item[\rm{(a)}]
$0\leq 1-\frac{\sharp{\cal M}_N}{M_N}\leq
\frac{1}{N^{2\alpha}}\frac{N}{M_N}$;
\item[\rm{(b)}] For all $f\in C^\infty(\T)$, and for all $k\in \N$,
 there exists a constant $C_{f,k}$ so that, for all
$0\leq t< (1-\epsilon)\log_2 N$ and for all $j\in {\cal M}_N$,
\begin{eqnarray*}
|\langle x_j^{(N)}, i| V_B^{-t}\opwo f V_B^t|x_j^{(N)}, i\rangle&
-& \langle x_j^{(N)}, i| \opwo {f\circ B^t}|x_j^{(N)},
i\rangle|\leq\\
&\,&\qquad\qquad C_{f,k}\bigl[\frac{1}{\sqrt{M_N}} + N^{\alpha -
\frac{\epsilon}{4}} +
N^{-k\frac{\epsilon}{2}}\bigr];
\end{eqnarray*}
\end{itemize}
{\rm ($ii$)} For each $f\in C^\infty(\T)$ there exist $C_{f_1}$,
$C_{f_2}$, $C_{f_3}$ such that $\forall x_0\in\tor$, $N,t>0$
$$
|\int_\tor dy \ g_{x_0,z}^{sc}(y,t)f(y) - \int_\tor dy\ f(y)| \leq
C_{f_1} \frac{\sqrt{N}}{2^t} + C_{f_2}\frac{2^t}{N} + C_{f_3}
e^{-\pi N\beta_-(z)/4}  \;.
$$
\end{proposition}
The second part of this proposition says that the semi-classically evolved
Husimi distribution equidistributes provided $\sqrt N << 2^t << N$. Comparing
to Theorem \ref{baker}, we conclude that its behaviour is  identical  to that
of the quantum mechanically evolved distribution on that time scale. We
unfortunately have no idea what happens at later times.
Analogously, for the position eigenstates we have:
\begin{proposition}
\label{q_baker_position}
{\rm [Position eigenstates]}   Let $N>0$,
$0<\epsilon<1$ and $0<\alpha<\epsilon/4$ be given. Let $f\in C^\infty(\T)$.
Then
\begin{itemize}
\item[(i)] There exists ${\cal M}_N\subset\{1,\ldots,N\}$ satisfying
$$
0\leq 1-\frac{\#{\cal M}_N}{N} \leq \frac{1}{N^{2\alpha}} \;;
$$
and for all $f\in C^\infty(\T)$, for all $k\in \N$, a constant $C_{f,k}>0$
such that for all $0\leq t< (1-\epsilon)\log_2N$ and for each $j\in{\cal
M}_N$
$$
|\langle e_j,\ V_B^{-t}\opwo fV_B^t e_j\rangle-f(\frac{2^tj}{N})|
\leq C_{f,k}\left(\frac{1}{N^{\epsilon/4-\alpha}}+\frac{1}{N^{\epsilon
k/2}}\right);
$$
\item [(ii)] There exists $C>0$ s.t., for $0\leq j\leq N-1$, $f\in
C^\infty(\T)$,
 $t,N>0$ we have
$$
|\int_{\tor}g_{j}^{sc}(x,t) f(x)\ dx -
\frac{1}{2}[f(\frac{2^tj}{N})+f(\frac{2^t(j+1)}{N})]| \leq
C \frac{\norm{f'}_\infty}{\sqrt{N}}  \;.
$$
\end{itemize}
\end{proposition}
The first part of this result is again readily understood in terms of
evolution of
the support of the Wigner function of the initial state $e_j$, which is the
vertical strip at $j/N$, of width $1/N$. The dynamics contracts this strip
vertically, stretches it horizontally to size $2^t/N$, and centers it at
$2^tj/N$.
As a result, one does not expect mixing to set in before times of order
$\log_2 N$, i.e. when $2^t/N\sim 1$; this is indeed confirmed by the above
result.  Comparing furthermore the first part of the proposition to the
second, one concludes again that the semi-classical evolution can not be
distinguished from the quantum-mechanical one up to times $2^t\ll N$.

\bigskip
\bigskip
\subsection{Proof of Propositions \ref{q_baker_cs} - \ref{q_baker_position}
and of Theorem \ref{baker}}
\label{qbaker2}
To prove part ($i$) of Proposition {\ref{q_baker_position}} as well as of
Proposition
\ref{q_baker_cs}, we first of all need the following  Egorov theorem, proven
in
\cite{dbde}.
Let's define, for each $t>0$ (compare this to (\ref{egorov})),
$$
E_t(f) = V_B^{-t} \opwo{f} V_B^t - \opwo{f\circ B^t}\;,
$$
and $E_t(m)=E_t(\chi_m)$.

\medskip
\begin{proposition}
\label{egorov_baker}\cite{dbde}
Let $\eta_N>0$ and $t_N>0$ such that $2^{t_N}\eta_N< N$. Then there exists
a subspace ${\cal G}_{\eta_N}(t_N)\subset{\cal H}_N$ such that
\begin{itemize}
\item[$i${\rm )}] ${\rm dim\ }{\cal G}_{\eta_N}(t_N)\geq N-2^{t_N}\eta_N$;

\item[$ii${\rm)}] for each $\psi\in{\cal G}_{\eta_N}(t_N)$, $|n|<\eta_N$ and
$0\leq t\leq t_N$
$$
E_t(n,0) \ \psi = 0 \;.
$$
\end{itemize}
\end{proposition}
\medskip
The Proposition asserts roughly that, provided one looks at times shorter
than
$\log_2 N$, and provided one restricts one's attention to a ``good" subspace
${\cal G}_{\eta_N}$, there is no error in the Egorov theorem for
trigonometric
polynomials of degree at most $\eta_N$.
The good subspace gets smaller as the time gets larger, but is non-trivial on
the
time-scale considered. Two obvious weaknesses of the above result are that it
does
not
describe the good space explicitly and that it deals only with functions of
$q$.
These are at the origin of the limitations of Theorem \ref{baker} pointed out
in
the
introduction.

The following corollary is an easy consequence of Proposition
\ref{egorov_baker}.
We write  $P_{{\cal B}_{\eta_N}}$ for the projector onto ${\cal B}_{\eta_N}$,
the
orthogonal
complement of ${\cal G}_{\eta_N}$.

\medskip
\begin{corollary}
\label{cor_egorov_baker}
Let $\{\phi_j\}_{j=1}^{M_N}$ be a family of
orthonormal vectors in ${\cal H}_N$. Let $\eta_N, t_N$ be as in Proposition
{\rm\ref{egorov_baker}}, $\delta_N>0$ and
$$
{\cal M}_N=\{1\leq j\leq M_N \ | \
\langle \phi_j|P_{{\cal B}_{\eta_N}} | \phi_j\rangle < \delta^2_N\} \;.
$$
Then
\begin{itemize}
\item[$i$)] $\#{\cal M}_N\geq M_N-2^{t_N}\eta_N/\delta_N^2$;
\item[$ii$)] for each $f\in C^\infty(\T)$ and $k\in\N$, there
exists $C_{f,k}>0$ such that, for each $0\leq t \leq t_N$ and $j\in{\cal
M}_N$,
$$
\norm{E_t(f)\phi_j} \leq
C_{f,k}(\delta_N+\eta_N^{-k})\;.
$$
\end{itemize}
\end{corollary}
We will always choose things in such a way that
$2^{t_N}\eta_N/\delta_N^2/M_N\to0, \delta_N\to0, \eta_N\to\infty$,
so that the Corollary asserts that the error is ``small" on ``many" $\phi_j$.

\noindent\dim ($i$) From Proposition \ref{egorov_baker} we have that
${\rm dim\ }{\cal B}_{\eta_N}\leq 2^{t_N}\eta_N$ and
$$
{\rm dim\ }{\cal B}_{\eta_N}={\rm Tr}P_{{\cal B}_{\eta_N}}\geq
\sum_{j\not\in{\cal M}_N}
\langle \phi_j|P_{{\cal B}_{\eta_N}}|\phi_j\rangle \geq
(M_N-\#{\cal M}_{N})\delta^2_N\;,
$$
from which the statement follows.

\smallskip
($ii$) Let $j\in{\cal M}_N$ and $f=\sum_n c_n\chi_{n0}\in C^\infty(\tor)$.
Then
$$
\norm{E_t(f)\phi_j} \leq \sum_{|n|<\eta_N} |f_n|\norm{E_t(n,0)\phi_j} +
\sum_{|n|\geq\eta_N} |f_n|\norm{E_t(n,0)\phi_j}.
$$
We then have, using Proposition \ref{egorov_baker},
$$\sum_{|n|<\eta_N} |f_n|\norm{E_t(n,0)\phi_j} =
\sum_{|n|<\eta_N} |f_n|\norm{ E_t(n,0)P_{ {\cal B}_{\eta_N} }\phi_j }
\leq \delta_N 2 \sum_{|n|<\eta_N} |f_n| \;,
$$
and
$$
\sum_{|n|\geq\eta_N} |f_n|\norm{E_t(n,0)\phi_j} \leq \frac{2}{\eta_N^k}
\sum_{|n|>\eta_N} |f_n| |n|^k\;.
$$
The result then follows with $C_{f,k}= 2 \sum_n |f_n||n|^k$. \fidi

The idea of the proofs of Propositions \ref{q_baker_cs} (i) and
\ref{q_baker_position}
(i) is
to apply the above corollary to the families $|x_i^{(N)},i\rangle$ and
$|e_j\rangle$,
respectively. Since the former family is not orthogonal, we will furthermore
use
the
following result
from linear algebra, which is proven through an application of the
Gram-Schmidt
orthogonalization procedure.
\begin{lemma}\label{gram} Let $u_1, u_2, \dots u_M \in {\cal H}_N$ be such
that,
for some
$\epsilon$ satisfying $8\epsilon M\leq 1$,
\begin{equation}
|\langle u_i, u_j\rangle -\delta_{ij}| \leq \epsilon. \label{aorth}
\end{equation}
Then the $u_i$ are linearly independent and there exists an orthonormal
basis\\
$w_1, w_2, \dots w_M$ of span$\{u_1, u_2, \dots u_M\}$ so that
\begin{equation}\label{d}
|| w_i -u_i|| \leq 7\epsilon \sqrt M\leq \frac{1}{\sqrt M}.
\end{equation}
\end{lemma}
\noindent \dim We will prove both statements at once using the Gram-Schmidt
orthogonalization procedure as follows. First, let $v_i=\frac{u_i}{||u_i||}$,
then
\begin{equation}\label{a}
||v_i - u_i||=|1-||u_i||\,|\leq \epsilon.
\end{equation}
Now define, for all $1\leq i\leq M$, $v'_i = v_i - \chi_i$, where $\chi_1=0$
and
$\chi_i$
is the orthogonal projection of $v_i$  onto the span of $v_1, \dots v_{i-1}$.
It is
then clear that, for all $1\leq i\not= j \leq M,\  \langle v'_i,
v'_j\rangle=0$ and
that
the span of $u_1, \dots u_{M}$ equals the span of $v'_1, \dots v'_{M}$.
We now first show that, for all $1\leq i \leq M$, $v'_i\not=0$ by showing
that
\begin{equation}\label{b}
||\chi_i||_{{\cal H}_N}\leq 3\epsilon \sqrt M.
\end{equation}
For $i=1$, (\ref{b}) is trivial. Consider then a fixed value of $i$ between
$2$ and
$M$ and  write
$
\chi_i = \sum_{j=1}^{i-1} \lambda^j_{i} v_j,
$
where the $\lambda^j_{i}$ are obtained by solving ($1\leq k < i$)
$$
\langle v_k, \chi_i\rangle = \langle v_k, v_i\rangle=
\sum_{j=1}^{i-1} \lambda^j_{i} \langle v_k, v_j\rangle.
$$
Introducing the matrices $[V_i]^{kj} = \langle v_k, v_j\rangle, \
 [\tau_i]^k=\langle v_k, v_i\rangle\  (1\leq k,j < i$), this can
be rewritten $\tau_{i}=V_i\lambda_i$. Consequently, since
$V_i^*=V_i$, we have
$$
\langle \chi_i,\chi_i\rangle=\lambda_i^*V_i\lambda_i =
\tau_i^*V_i^{-1}V_i V_i^{-1}\tau_i=\tau_i^*V_i^{-1} \tau_i,
$$
and
$$
||\chi_i||_{{\cal H}_N}^2\leq
||V_i^{-1}||\left(\sum_{k=1}^{i-1}|\langle v_k, v_i\rangle|^2
\right).
$$
To estimate $||V^{-1}_i||$, note that $V_i=I+\epsilon S_i$, where
$S_i$ is an off-diagonal self-adjoint matrix with matrix elements
$(k\not= j)$
$$
|S^{kj}_i|=\frac{1}{\epsilon}|\langle v_k, v_j\rangle| =
\frac{|\langle u_k, u_j\rangle|} {||u_k||\ ||u_j||\ \epsilon}\leq
\frac{2\epsilon}{\epsilon}.
$$
Consequently,
$$
||S_i||=\sup_{||a||_{\R^{(i-1)}}=1}|\langle a, S
a\rangle_{\R^{(i-1)}}|\leq 2\sup_{||a||_{\R^{(i-1)}}=1}
\sum_{k,l=1}^{i-1}|a_k|\ |a_l|\leq 2M.
$$
As a result
$$
||V^{-1}_i||\leq \sum_{r=0}^\infty \epsilon^r ||S_i||^r
\leq\frac{1}{1-\epsilon||S_i||}\leq \frac{4}{3}
$$
and consequently
$
||\chi_i||^2_{{\cal H}_N}\leq (4/3)4\epsilon^2 M\leq 6\epsilon^2 M,
$
from which (\ref{b}) follows.

Since all $v'_i\not= 0$, we can define $w_i=v'_i/||v'_i||$
with $\langle w_i, w_j\rangle=\delta_{ij}$.  Since furthermore
$$
\hbox{span}\{u_1, u_2, \dots u_M\}=\hbox{span}\{v'_1, v'_2, \dots v'_M\}=
\hbox{span}\{w_1, w_2, \dots w_M\}
$$
it follows that the $u_i$ are linearly independent. It remains to show
(\ref{d}).
To that end,
compute, using (\ref{a}),
$$
||w_i - u_i||\leq ||w_i-v'_i|| + ||v'_i -v_i|| + ||v_i-u_i||\leq
|1-||v'_i||\ | + ||\chi_i|| + \epsilon.
$$
Since $|1-||v'_i||\ |=|\ ||v_i|| - ||v'_i||\ |\leq ||v'_i
-v_i||=||\chi_i||$, the result follows from (\ref{b}).\fidi

\noindent {\sl Proof of Proposition  \ref{q_baker_cs}.}
($i$) We first use Lemma \ref{gram} and
(\ref{cohstateoverlap})-(\ref{mncondition}) to assert the
existence of an orthonormal set $\phi_j, 1\leq j_1, j_2\leq
\sqrt{M_N}$ such that
$$
||\phi_j - |x_j^{(N)}, i\rangle||\leq 7\sqrt{M_N} C
e^{-\frac{\pi}{2}\frac{N}{M_N}} \leq \frac{1}{\sqrt{M_N}} \;.
$$
We then apply Corollary \ref{cor_egorov_baker} with
$\eta_N=N^{\frac{\epsilon}{2}}, \delta_N=
N^{\alpha-\frac{\epsilon}{4}}$ and $t_N=(1-\epsilon)\log_2N$, so
that
$$
\frac{\sharp{\cal M}_N}{M_N}\geq 1-\frac{N^{1-\epsilon}}{M_N} N^{\epsilon/2}
N^{-2\alpha
+\frac{\epsilon}{2}}
$$
and for all $j\in{\cal M}_N, \ 0\leq t\leq t_N$
$$
||E_t(f)\phi_j||\leq C_{f,k}(N^{\alpha-\frac{\epsilon}{4}}+
N^{-k\frac{\epsilon}{2}}).
$$
Consequently,
\begin{eqnarray*}
|\langle x_j^{(N)}, i| V_B^{-t}\opwo{f} V_B^t|x_j^{(N)},i\rangle
&-& \langle x_j^{(N)}, i| \opwo{f\circ B^t} |x_j^{(N)},i\rangle|
\cr
& &\leq \||x_j^{(N)},i\rangle\|
\|E_t(f)|x_j^{(N)},i\rangle\|\cr
& & \leq C (\|E_t(f)(\phi_j-|x_j^{(N)},i\rangle)\|+
\|E_t(f)\phi_j\|) \cr
& &\leq C_{f,k}\bigl[\frac{1}{\sqrt{M_N}}+
N^{\alpha-\frac{\epsilon}{4}} +N^{-k\frac{\epsilon}{2}}\bigr],
\end{eqnarray*}
which is the desired result.

($ii$) Since the Baker map is linear on $f\in C^\infty(\T)$, this  is
a direct consequence of the proof of Proposition \ref{direct_proof}. \fidi

\noindent {\sl Proof of Theorem \ref{baker}.}
With the notations of Proposition
\ref{q_baker_cs}, let $x_0\in\tor$ and consider
$$
d_N(x_0)= \min\{|x_j^{(N)}-x_0|\, |\, j\in {\cal M}_N\}.
$$
Since $M_N -\sharp{\cal M}_N \leq \frac{N}{N^{2\alpha}}$, it is clear that
$$
d_N(x_0)\leq \sqrt{\frac{N}{M_N}}\frac{1}{N^{\alpha}}\;.
$$
Let's choose $\alpha > \epsilon/5$; then the sequence $M_N$
defined after (\ref{cs_diseq}) is such that
$N/M_N<N^{2(\alpha-\epsilon/5)}$. Consequently $\forall N, \exists
j_N\in {\cal M}_N$ such that $||x_0-x_{j_N}^{(N)}||\leq N^{-\epsilon/5}$. We
write $x_N=x_{j_N}^{(N)}$.

Let $t<\frac{1-\epsilon}{2}\log_2 N$ and consider the following
inequality:
$$
|\langle x_N,i| V_B^{-t} Op^{AW}_{0,z}(f) V_B^t|x_N,i\rangle -
f(2^tq_N)| \leq C || Op^{W}_0(f)- Op^{AW}_{0,z}(f)|| +
$$
$$
|\langle x_N,i| E_t(f)|x_N,i\rangle| + |\langle x_N,i|
Op^W_0(f\circ B^t) |x_N,i\rangle - f(2^tq_N)| \;.
$$

The result then comes from Proposition \ref{q_baker_cs}, the
general estimate $||\opwo{f}-\opaw{f}||\leq C_f/N$, Proposition
\ref{plane_limit}(i) with $A=B_o$ and a straightforward adaptation of
Proposition \ref{torus_limit} to the case $A=B_o$ and $f=f(q)$. In
the same way we obtain the result for the mixing regime. \fidi

\noindent {\sl Proof of Proposition {\ref{q_baker_position}}}.
($i$)  With respect to Corollary \ref{cor_egorov_baker} let $M_N=N$,
$\phi_j=e_j$,
$\eta_N=N^{\epsilon/2}$, $\delta_N=N^{\alpha-\epsilon/4}$ and
$t_N\leq(1-\epsilon)
\log_2 N$.
Then, we have that $\#{\cal M}_N/N\geq 1 - 1/N^{2\alpha}$ and
$$
|\langle e_j|E_t(f)| e_j\rangle| \leq C_{f,k}
\left(N^{\alpha-\epsilon/4}+N^{-k\epsilon/2}\right)\;.
$$
Finally, by a simple computation we find that, if $\partial_p f=0$,
$$
\langle e_j | \opwo{f\circ B^t} |e_j\rangle = f(\frac{2^tj}{N})
\;.
$$

\smallskip
{\rm ($ii$)} Following the method of Proposition \ref{fourier_gen}, we write
the Fourier series in $\kappa$ and obtain for $\kappa = 0$
$$
\int_\tor g^{sc}_j(x,0) f(x)dx = \sum_{\ell\in\Z^2}
\langle \epsilon_j| \opaw f |U(\ell)\epsilon_j\rangle
$$
$$
\qquad = N^2\sum_{\ell\in\Z^2}\int_{{\bf R}^2} dx f(x)
\int_{\frac{j}{N}}^{\frac{j+1}{N}}ds\ \eta_{x,z}(s)
\int_{\frac{j}{N}-\ell_1}^{\frac{j+1}{N}-\ell_1}dw\ \eta_{x,z}^*(w)e^{i2\pi
N\ell_2w}\;,
$$
where we used the definition of $\epsilon_j$ (Lemma \ref{lem_posvec}) and the
explicit action of $U(\ell)$. Here $\eta_{x,z}$ are the coherent states
defined in (\ref{cs}). Introducing furthermore $\phi_{q,z}(s)= \exp(i\pi N
pq)
\exp(2i\pi Nps)\eta_{x,z}(s)$ and recalling that $f$ depends only on $q$ one
obtains, as result of the integration in $p$, $\delta(s-w)$; after
integrating over $w$ this yields
$$
\int_\tor g^{sc}_j(x,0) f(x) dx = N \sum_{\ell_2}
\int dq f(q) \int_{\frac{j}{N}}^{\frac{j+1}{N}}ds |\phi_{q,z}(s)|^2e^{i2\pi N
\ell_2 s} \;.
$$
If we introduce the characteristic function $\chi_j$ of the interval
$[j,j+1]$ we
see that
\begin{eqnarray*}
\int_\tor g^{sc}_j(x,0) f(x) dx &=&
    \int dq\ f(q) \sum_{\ell\in\Z}\int_\R|\phi_{q,z}(\frac{s}{N})|^2
\chi_j(s)
e^{i2\pi \ell s}\ ds\cr
&=& \frac{1}{2}\int dq\ f(q) (\lim_{\epsilon\rightarrow
0^+}+\lim_{\epsilon\rightarrow 0^-})
\sum_\ell |\phi_{q,z}(\frac{\ell+\epsilon}{N})|^2\chi_j(\ell+\epsilon) \cr
&=& \frac{1}{2}\int dq\ f(q)
\left[|\phi_{q,z}(\frac{j}{N})|^2 + |\phi_{q,z}(\frac{j+1}{N})|^2\right]\;,
\end{eqnarray*}
where we used in the second line the convergence properties of the Fourier
series
of
$|\phi_{q,z}(s/N)|^2\chi_j(s)$. If $z=i\omega$, using the explicit form of
$\phi_{q,z}$
we easily obtain the following estimate with $C>0$ an universal constant
$$
|\int dx g_j^{sc}(x,0)f(x)-\frac{1}{2}( f(\frac{j}{N})+f(\frac{j+1}{N}))|
\leq C \frac{1}{\sqrt{\omega N}} \norm{\partial_q f}_\infty \;.
$$
To obtain the final result concerning the $t$-dependence, observe that
it is enough to change $f\rightarrow f\circ B^t$ and $\omega\rightarrow
4^t\omega$.
\fidi

\enddocument